\documentclass[sigconf,balance=false]{acmart}
\usepackage{popets}
\usepackage{cleveref}

\usepackage{booktabs,xcolor,pifont, array}
\usepackage{algorithm}
\usepackage{algpseudocode}
\usepackage{tabularx}
\usepackage{makecell}
\usepackage{enumitem}
\newcommand{\greenx}{\textcolor{green!60!black}{\ding{55}}} 
\newcommand{\redcheck}{\textcolor{red!70!black}{\ding{51}}}
\newcolumntype{C}[1]{>{\centering\arraybackslash}m{#1}}
\usepackage{tikz}
\usepackage{adjustbox} 
\usetikzlibrary{positioning, arrows.meta, calc, backgrounds, fit, shapes}
\usepackage{listings}
\usepackage{booktabs}  
\usepackage{hyperref}  
\usepackage{makecell}
\usepackage[normalem]{ulem}

\newif\iftrackchanges
\trackchangestrue 

\lstdefinestyle{mypython}{
    language=Python,
    basicstyle=\ttfamily\small, 
    keywordstyle=\color{blue!70!black}\bfseries,
    stringstyle=\color{red!70!black},   
    commentstyle=\color{green!50!black}\itshape,
    morecomment=[l][\color{purple!60!black}]{\#}, 
    numbers=left,         
    numberstyle=\tiny\color{gray}, 
    frame=tb,              
    showstringspaces=false,     
    breaklines=true,        
    captionpos=b         
}

\lstdefinestyle{terminal}{
    basicstyle=\ttfamily\small,
    backgroundcolor=\color{black!5},
    frame=single,
    rulecolor=\color{black!30},
    numbers=none,
    breaklines=true,
    showstringspaces=false,
    columns=fullflexible,
}

\usepackage[colorinlistoftodos,textsize=footnotesize]{todonotes}
\presetkeys{todonotes}{inline}{}

\setcopyright{popets}
\copyrightyear{YYYY}

\acmYear{YYYY}
\acmVolume{YYYY}
\acmNumber{X}
\acmDOI{XXXXXXX.XXXXXXX}
\acmISBN{}
\acmConference{Proceedings on Privacy Enhancing Technologies}
\newtheorem{assumption}{Assumption}[section]
\newtheorem*{assumption*}{Assumption}

\begin{document}

\title[Privacy in Theory, Bugs in Practice: A Grey-Box Auditing Framework of Differential Privacy Libraries]{Privacy in Theory, Bugs in Practice: \\ Grey-Box Auditing of Differential Privacy Libraries}

\author{Tudor Cebere}
\affiliation{%
  \institution{PreMeDICaL team, Inria, Idesp, Inserm, Université de Montpellier}
  \city{}
  \state{}
  \country{}}
\email{tudor.cebere@inria.fr}

\author{David Erb}
\affiliation{%
  \institution{Technical University of Munich \& Oblivious}
  \city{}
  \country{}}
\email{david.erb@tum.de}

\author{Damien Desfontaines}
\affiliation{%
  \institution{Hiding Nemo}
  \city{}
  \country{}
}
\email{damien@hiding-nemo.ch}

\author{Aurélien Bellet}
\affiliation{%
\institution{PreMeDICaL team, Inria, Idesp, Inserm, Université de Montpellier}
 \city{}
 \state{}
 \country{}}
\email{aurelien.bellet@inria.fr}

\author{Jack Fitzsimons}
\affiliation{%
  \institution{Oblivious}
  \city{}
  \state{}
  \country{}}
\email{jack@oblivious.com}

\newcommand{\methodtrusted}{Re:cord-play}
\newcommand{\methoduntrusted}{Re:cord-play-sample}

\renewcommand{\shortauthors}{Cebere et al.}

\begin{abstract}
Differential privacy (DP) implementations are notoriously prone to errors, with subtle bugs frequently invalidating theoretical guarantees. Existing verification methods are often impractical: formal tools are too restrictive, while black-box statistical auditing is intractable for complex pipelines and fails to pinpoint the source of the bug.
This paper introduces \methodtrusted, a ``gray-box'' auditing paradigm that inspects the internal state of DP algorithms. By running an instrumented algorithm on neighboring datasets with identical randomness, \methodtrusted{} directly checks for data-dependent control flow and provides concrete falsification of sensitivity violations by comparing declared sensitivity against the empirically measured distance between internal inputs. We generalize this to \methoduntrusted, a full statistical audit that isolates and tests each component, including untrusted ones.
We show that our novel testing approach is both effective and necessary by auditing 12 open-source libraries, including SmartNoise SDK, Opacus, and Diffprivlib, and uncovering 13 privacy violations that impact their theoretical guarantees.
We release our framework as an open-source Python package\footnote{\url{https://github.com/ObliviousAI/dp-recorder}}, thereby making it easy for DP developers to integrate effective, computationally inexpensive, and seamless privacy testing as part of their software development lifecycle.
\end{abstract}


\maketitle

\section{Introduction}
\label{sec:intro}
The proliferation of large-scale data collection and analysis has made protecting individuals' privacy a key challenge for society and technology. 
In response, differential privacy (DP) has emerged as the de facto standard for private data sharing, offering a rigorous, provable privacy guarantee~\cite{dwork_2016}. A randomized mechanism is said to be differentially private if its output distribution remains statistically indistinguishable when applied to any pair of neighboring datasets, i.e., datasets differing by the data of a single individual. 
This guarantee ensures that the presence or absence of any individual's data in a dataset has a mathematically bounded impact on the outcome, protecting them from privacy attacks based on released results. 
In most DP mechanisms, this influence is captured by the \emph{sensitivity} of the underlying query, i.e., the maximum change in the query's output when the data of a single individual is modified. DP also supports building complex algorithms by composing basic \emph{privacy primitives} (e.g., the Laplace, Gaussian, and exponential mechanisms) and applying data-independent post-processing to their outputs.
These properties have led to the growing adoption of DP by technology companies and public institutions \cite{aktay2020google, zhang2023differentially, johnson_towards_2018, abowd_us_2018}\footnote{For a comprehensive list, see the Differential Privacy Deployments Registry \cite{howarth_communitydriven_2025}.}, making it a critical component in the modern data ecosystem.

\newsavebox{\auditdiagrambox}

\begin{lrbox}{\auditdiagrambox}
    \begin{tikzpicture}[
        node distance=1.5cm and 4cm,
        base/.style = {rectangle, rounded corners, minimum width=3.5cm, minimum height=1.2cm, align=center, font=\sffamily},
        public/.style = {base, fill=blue!10, draw=blue!60!black, thick, minimum width=2.8cm, minimum height=0.8cm}, 
        private/.style = {base, fill=red!10, draw=red!60!black, line width=1.5pt},
        header/.style = {font=\sffamily\bfseries\large, align=center},
        arrow/.style = {-{Latex[length=3mm]}, thick, darkgray},
        labeltext/.style = {font=\sffamily\small, align=center}
    ]
        \node[header, text=blue!40!black] (h1) {PHASE 1: RECORD\\(Dataset $D$)};
        \node[public, below=0.6cm of h1] (pre1) {Pre-processing};
        \node[labeltext, below=0.6cm of pre1, text=red!70!black] (in1) {Input: $q(D)$};
        \node[private, below=0.4cm of in1] (mech1) {\textbf{Mechanism Call}\\(e.g., Laplace)};
        \node[labeltext, below=0.4cm of mech1, text=green!40!black] (out1) {Output: $y = 5.2$};
        \node[public, below=0.6cm of out1] (post1) {Post-processing}; 
        \node[public, right=of pre1] (pre2) {Pre-processing}; 

        \node[header, text=blue!40!black, above=0.6cm of pre2] (h2) {PHASE 2: REPLAY\\(Dataset $D'$)};
        \node[public, right=of pre1] (pre2) {Pre-processing};
        \node[labeltext, below=0.6cm of pre2, text=red!70!black] (in2) {Input: $q(D')$};
        \node[private, below=0.4cm of in2] (mech2) {\textbf{Mechanism Call}\\(Replay Mode)};
        \node[labeltext, below=0.4cm of mech2, text=green!40!black] (out2) {Forced Output: $y = 5.2$};
        \node[public, right=of post1] (post2) {Post-processing}; 

        \draw[arrow] (h1) -- (pre1);
        \draw[arrow] (pre1) -- (in1);
        \draw[arrow] (in1) -- (mech1);
        \draw[arrow] (mech1) -- (out1);
        \draw[arrow] (out1) -- (post1);

        \draw[arrow] (h2) -- (pre2);
        \draw[arrow] (pre2) -- (in2);
        \draw[arrow] (in2) -- (mech2);
        \draw[arrow] (mech2) -- (out2);
        \draw[arrow] (out2) -- (post2);

        \draw[{Latex}-{Latex}, dotted, line width=1.5pt, color=violet] 
            (pre1.east) -- node[midway, fill=white, font=\sffamily\bfseries\footnotesize, align=center] {0. CHECK INVARIANCE} 
            (pre2.west);

        \draw[->, dashed, line width=1.5pt, color=green!60!black] 
            ($(out1.east)+(0.1,0)$) -- node[midway, above, font=\sffamily\bfseries\footnotesize, fill=white] {2. FREEZE OUTPUT} 
            ($(out2.west)-(0.1,0)$);

        \draw[{Latex}-{Latex}, line width=1.5pt, color=red!60!black] 
            ($(in1.east)+(0.1,0)$) -- node[midway, above, font=\sffamily\bfseries\footnotesize, fill=white] {1. CHECK SENSITIVITY\\$|q(D) - q(D')| \le \Delta$?} 
            ($(in2.west)-(0.1,0)$);

        \draw[{Latex}-{Latex}, dotted, line width=1.5pt, color=violet] 
            (post1.east) -- node[midway, fill=white, font=\sffamily\bfseries\footnotesize, align=center] {3. CHECK INVARIANCE} 
            (post2.west);

        \begin{scope}[on background layer]
            \path (h1.north west)+(-0,0) node (a) {};
            \path (post2.south east)+(0.23,0) node (b) {};
            \node[fit=(a)(b), draw=gray!20, fill=gray!5, rounded corners=10pt] {};
        \end{scope}
    \end{tikzpicture}
\end{lrbox}

\begin{figure}[t]
    \centering
    \resizebox{\linewidth}{!}{\usebox{\auditdiagrambox}}
    \caption{Summary of \methodtrusted\ highlighting the Record Phase (Dataset $D$) and Replay Phase (Dataset $D'$). }
    \label{fig:record_replay}
\end{figure}

Unfortunately, the conceptual elegance of DP contrasts with the practical difficulty of implementing it correctly.
DP guarantees are notoriously susceptible to subtle implementation flaws that can invalidate the claimed protection.
These bugs arise from sources including incorrect sensitivity calculations \cite{tramer_debugging_2022}, data-dependent pre- or post-processing \cite{ganev2025importance, annamalai2024you}, mishandled privacy budget composition \cite{lyu_understanding_2017}, insecure randomness \cite{safavi-naini_differential_2012}, or floating-point vulnerabilities \cite{haney2022precision, mironov2012, jin2022we}.
This creates a gap between the stated privacy guarantees and the actual protection provided by their implementation.

To address this gap between theoretical promises and practical guarantees, the research community has increasingly turned to \emph{distributional auditing} \cite{annamalai2025hitchhiker}.
This statistical approach aims to empirically falsify privacy guarantees by distinguishing the output distributions of a mechanism on neighboring datasets.
The majority of distributional auditing frameworks operate as \emph{black-box} auditors, assessing privacy properties based solely on input-output behavior.
This approach can detect privacy violations regardless of their cause, from high-level algorithmic mistakes to low-level implementation bugs, thus providing a robust assessment of a system.

However, such techniques are limited when auditing modern DP algorithms, which are rarely simple primitives but rather complex pipelines that chain heterogeneous components. A typical workflow involves preprocessing (like clipping), adaptive composition of multiple mechanisms, and post-processing of their outputs. Black-box auditing of such pipelines presents two challenges. First, it is often statistically intractable: the final output distribution is the high-dimensional result of a complex process, requiring a prohibitive number of samples to distinguish. Second, it is uninformative: even if a black-box test detects a violation, it flags the entire system as faulty without pinpointing the faulty component.

This leaves developers with few practical options. Formal verification tools \cite{gaboardi2013linear}, while powerful, are typically restrictive, requiring entire pipelines to be rewritten in specialized, provable domain-specific languages (DSLs). This is a barrier to entry for the vast ecosystem of Python or C++ libraries.
Thus, a practical gap remains: we lack an automated tool that can pinpoint where and why an implementation fails without the cost of end-to-end auditing.

Our work is motivated by the need for a ``privacy debugger'' integrated into the software development lifecycle. Rather than relying on manual verification, we propose a solution to detect implementation bugs in Continuous Integration and Delivery (CI/CD) pipelines. We aim to provide a tool as effortless as a standard unit test, yet capable of evaluating highly randomized algorithms.

\paragraph{Contribution} This paper bridges the gap between theoretical privacy guarantees and practical correctness of implementations by introducing a new \emph{gray-box
auditing} paradigm. Unlike black-box methods that analyze high-dimensional output distributions, our approach assumes access to the underlying implementation and inspects the internal state of the algorithm mechanisms and the interfaces between its components. Our core insight is that DP implementations are not monolithic: they interleave data-independent logic with calls to data-dependent privacy primitives. We leverage the fact that most privacy violations that arise fall into one of the following categories: (i) data-dependent leakage into data-independent code, (ii) sensitivity miscalculations, (iii) noise miscalibration, and (iv) flawed privacy primitive implementations. Our approach generates concrete, actionable reports that help practitioners identify and resolve common errors.
Our contributions can be summarized as follows:

\begin{enumerate}
    \item \textbf{Deterministic Verification (\methodtrusted):} We leverage a novel record-and-replay instrumentation that benefits from the separation between private primitives and the rest of the processing logic (see ~\Cref{fig:record_replay}). By \emph{freezing} the outputs of privacy primitives during execution on the neighboring dataset, we check whether code assumed to be data-independent indeed remains \emph{invariant} across datasets.
    This allows us to catch leaks of type (i) deterministically: if intermediate variables diverge despite identical mechanism outputs, data has leaked into logic assumed to be data-independent. Simultaneously, it flags sensitivity errors of type (ii) by comparing the declared privacy parameters against the empirically measured distance between the internal inputs fed to each primitive. Additionally, if we assume \emph{vetted, trusted, and correctly implemented} privacy primitives, we can catch errors of type (iii) as well.
    \item \textbf{Component-Wise Audit (\methoduntrusted):} We extend our framework to handle untrusted or custom primitives via \methoduntrusted. By isolating privacy primitives, we perform statistical tests on them rather than the entire pipeline, making the auditing problem computationally cheaper and more tractable. Applying ideas from black-box auditing within our framework, we propose to leverage Privacy Loss Distributions (PLDs) to obtain precise privacy loss estimates for each privacy primitive. This approach allows us to detect issues of type (iii) and (iv), while enabling the composition of both analytically known PLDs and empirically estimated PLDs from arbitrary mechanisms to derive end-to-end privacy bounds.
    \item \textbf{Practical Evaluation:} We demonstrate the practical value of our framework by auditing popular open-source DP libraries, including SmartNoise SDK, Synthcity, Opacus, and Diffprivlib, in which we \emph{uncover several previously unknown privacy violations}. These findings range from subtle sensitivity miscalculations in data preprocessing to data-dependent logic flaws. This validates our approach as a practical, effective method for finding real-world bugs that elude current testing and verification techniques. We release our framework as an open-source package, lowering the barrier to test DP code and encouraging practitioners to adopt our effective and computationally inexpensive tool as a component of their software development lifecycle.
\end{enumerate}

The rest of this paper is organized as follows. We discuss background and related work in ~\Cref{sec:background}. In ~\Cref{sec:motivation}, we introduce our auditing techniques, \methodtrusted and \methoduntrusted. We present our evaluation setup and the real-world bugs we uncovered in ~\Cref{sec:evaluation}. We provide practical insights and recommendations for authors of DP software, outline the limitations of our approach, and discuss future work in ~\Cref{sec:suggestionsforpractitioners}, and then conclude.

\section{Background \& Related Work}
\label{sec:background}
\subsection{Differential Privacy}

Differential Privacy (DP) \cite{dwork_2016} has become the standard for private data analysis, valued for its rigorous and mathematically provable guarantees. It enables the release of useful aggregate statistics from a dataset while limiting the disclosure of information about any single individual within it. A mechanism's privacy is quantified by how indistinguishable its output distributions are when run on any two neighboring datasets (i.e., datasets that differ by a single record). More formally, a mechanism is considered differentially private if it satisfies the following definition:

\begin{definition}
A randomized mechanism $\mathcal{M}:\mathcal{D}\rightarrow\mathcal{X}$ satisfies $(\varepsilon, \delta)$-approximate DP if for every pair of neighboring datasets $D, D'$, and every subset $A\subseteq\mathcal{X}$ of the output space, it holds that:
\begin{equation*}
     P(\mathcal{M}(D)\in A)\le e^{\varepsilon}P(\mathcal{M}(D')\in A)+\delta.
\end{equation*}
\end{definition}

Here, $\varepsilon$ represents the privacy loss parameter, while $\delta$ should be a small constant, see \cite{vaudenay_our_2006, desfontaines2019sok} for more details. Mechanisms with $\delta=0$ satisfy pure DP.

The core challenge in implementing any DP mechanism $\mathcal{M}$ is to correctly calibrate the amount of noise, which is directly determined by the mechanism's \emph{sensitivity}. The global sensitivity of a query $f: \mathcal{D} \rightarrow \mathbb{R}^k$ measures the maximum possible change in $f$'s output when applied to any two neighboring datasets, $D$ and $D'$ \cite{dwork_2016}. The specific definition of sensitivity depends on the $L_p$ norm used to measure this change: $\Delta_p f = \max_{D, D'} \|f(D) - f(D')\|_p$.

While our approach is general, we anchor our exposition in three canonical privacy primitives defined by their corresponding sensitivity: (i) the \emph{Laplace Mechanism} achieves pure $\varepsilon$-DP by adding Laplace-distributed noise scaled to the $L_1$-sensitivity of the query; (ii) the Gaussian Mechanism guarantees $(\varepsilon, \delta)$-DP by adding Gaussian noise, with a scale $\sigma$ calibrated to the $L_2$-sensitivity; (iii) the Exponential Mechanism \cite{mcsherry_mechanism_2007} serves as a general-purpose tool for discrete outputs, providing $\varepsilon$-DP by selecting an item $r \in \mathcal{R}$ based on a quality score $q$ with probability proportional to $\exp\left(\frac{\varepsilon q(D, r)}{2 \Delta_{\infty} q}\right) $, where $\Delta_{\infty} q$ is the $L_{\infty}$-sensitivity of the quality function.

Two fundamental properties make DP a powerful and practical framework. First, immunity to \emph{post-processing}: an analyst cannot weaken the privacy guarantee by performing arbitrary data-independent computations on the output of a DP mechanism. This ensures that once data is released privately, it remains private regardless of subsequent analysis. Second, \emph{composition} allows for managing a ``privacy budget'' across multiple analyses. Composition theorems \cite{dwork_algorithmic_2013, kairouz2015composition, vadhan_concurrent_2023, kushilevitz_complexity_2016} state that the total privacy loss from releasing results of multiple DP mechanisms on the same underlying data is a function of the privacy losses of the individual mechanisms. This allows for the construction of complex, multi-step private algorithms from simpler DP building blocks. To track this cumulative privacy loss, \emph{privacy accountants} \cite{abadi_deep_2016} are used, providing tighter bounds than basic composition theorems \cite{dwork_2016, dwork_algorithmic_2013}. This is often done by converting $(\varepsilon, \delta)$ into a representation for which composition is tight, such as Rényi DP (RDP) \cite{mironov_renyi_2017} or \emph{PLDs}~\cite{gopi_numerical_2021, doroshenko2022connect}, which can be composed more precisely before being converted back to the final $(\varepsilon, \delta)$ guarantee.

\subsection{Membership Inference Attacks and Auditing}

Membership Inference Attacks (MIAs) \cite{homer_resolving_2008, yeom_privacy_2018, shokri2017membership, ye_enhanced_2022, mahloujifar_optimal_2022, humphries_investigating_2023, annamalai2024you} determine whether a specific individual’s data was included in the input dataset of a function, based solely on its observed output. Typically, these attacks operate by comparing the outputs of a mechanism on datasets that either include or exclude a target record. If the outputs are sufficiently different, the adversary can infer the presence of the record with better than random accuracy. The strength of an MIA is typically evaluated by its advantage over random guessing, as visualized by the Receiver Operating Characteristic (ROC) curve. This analysis often focuses on the attack's True Positive Rate (TPR) at low False Positive Rates (FPRs) (e.g., $1\%$ or $0.1\%$), as this demonstrates the attack's practical effectiveness \cite{rezaei_difficulty_2021, carlini_membership_2022}.

In practice, constructing such attacks ranges in difficulty.  For simple algorithms,  one can often design linear or threshold-based classifiers that succeed \cite{dong2022gaussian, nasr2023tight}. 
For more complex algorithms, shadow-modeling~\cite{shokri2017membership} techniques are employed, which train auxiliary models to simulate the target mechanism and guide inference. These methods, however, are computationally demanding and algorithm-specific, requiring careful design and significant resources.

MIAs are particularly relevant when auditing differentially private algorithms \cite{jagielski2020auditingdifferentiallyprivatemachine, nasr_adversary_2021, zanella2023bayesian, nasr2023tight, annamalai2024s, chadha2024auditing, steinke_privacy_2023, mahloujifar2024auditing, cebere2024tighter, koskela2025auditing} because the DP definition is interpreted as a worst-case bound against membership inference. A mechanism satisfying $(\varepsilon,\delta)$-DP ensures that, for any record, the success probability of a membership inference attack for all FPRs \cite{wasserman_statistical_2009, kairouz2015composition, dong2022gaussian}, imposing a constraint on the ROC curve on the success probability of MIAs of the form $\text{TPR} \leq e^\epsilon \text{FPR} + \delta$. Thus, the formal DP guarantee provides a direct upper bound on the information leakage that MIAs can exploit. This relationship enables auditing privacy claims by inverting it: the empirical performance of an MIA directly yields a lower bound on the privacy parameters of an algorithm. If an adversary achieves a TPR/FPR pair that violates the theoretical bound, the claimed $(\varepsilon, \delta)$ is falsified.

We define a \emph{distributional auditing} pipeline (see \Cref{alg:audit-pipeline} in the Appendix) as a statistical framework that estimates the privacy loss of a mechanism by analyzing its output samples. While typically deployed as a \emph{black-box} (where the adversary observes only final outputs), the underlying statistical methodology remains the same regardless of access. It consists of two main components:
\begin{enumerate}
    \item \textbf{The Adversary ($\mathcal{A}$):} This component defines the attack strategy. It typically involves two subroutines: (i) a \texttt{Rank} function assigning a confidence score that the output was generated by $D$ or $D'$, and (ii) a \texttt{Reject} (e.g., a threshold) to classify the output as belonging to $\mathcal{M}(D)$ or $\mathcal{M}(D')$.
    \item \textbf{The Auditing Scheme:} This component analyzes the adversary's performance by running the mechanism many times on $D$ and $D'$. It computes the Type I error rate ($\alpha$) and Type II error rate ($\beta$) and uses these statistics to compute a high-probability lower bound on the $(\hat{\varepsilon}, \delta)$ privacy loss.
\end{enumerate}

\subsection{Related Work}
\label{sec:relatedwork}

\paragraph{Distributional Auditing.} A significant amount of the literature has been devoted to auditing DP machine learning (ML) algorithms. For a comprehensive summary, refer to \citet{annamalai2025hitchhiker}. Initially investigated by \citet{jagielski2020auditingdifferentiallyprivatemachine}, auditing DP-ML is constrained by the fact that the underlying mechanism is computationally intensive and typically involves subsampling \cite{kasiviswanathan2011can, DBLP:journals/corr/abs-1807-01647}. Recent research aims to improve the sample efficiency of these audits \cite{zanella2023bayesian, steinke2023privacy, mahloujifar2024auditing, koskela2025auditing}, audit various DP variants \cite{nasr2023tight, chadha2024auditing}, or improve accounting in specific threat models \cite{cebere2024tighter, annamalai2024s, annamalai2024shuffle}. We draw on standard techniques from this line of work in \Cref{sec:distributional-audit}, applying them to simple privacy primitives with unknown structures.

\paragraph{Optimization and Counterexample Search.} A central challenge in distributional auditing is identifying worst-case inputs, or counterexamples, yielding the tightest privacy lower bounds (see e.g. the hierarchy in \citet{nasr_adversary_2021}). \citet{10.1145/3243734.3243863} introduced DP-Finder, which formulates the search for privacy violations as an optimization problem and uses a smooth surrogate objective to discover high-violation counterexamples. Other approaches include StatDP \citep{ding2018detecting}, a semi-black-box tool using hypothesis testing, and DP-Sniper \citep{9519405}, which trains classifiers to distinguish neighboring outputs. Our framework complements these techniques: while these methods excel at optimizing for inputs triggering violations, our work focuses on pinpointing the bugs causing the harm.

\paragraph{Proofs of Privacy.} Another approach to auditing leverages Zero-Knowledge Proofs (ZKPs) \cite{10.1145/22145.22178}. While substantial work \cite{narayan_verifiable_2015, biswas_interactive_2023, shamsabadi_confidential-dpproof_2024, bellclark2025verity} has been done to certify DP compliance without data access, these proofs are limited in scope. As noted, they strictly demonstrate that the mechanism implemented in practice matches the intended algorithmic description. They do not validate the algorithm's actual behavior on the underlying data, nor can they verify whether the claimed privacy guarantee associated with the intended algorithm is theoretically valid \cite{nasr_adversary_2021, tramer_debugging_2022}. Therefore, this line of work addresses execution integrity and complements our approach.

\paragraph{Sensitivity-tracking and domain-specific languages (DSLs)} A robust method for preventing privacy violations is the use of libraries that allow only a subset of vetted functions. By composing these known transformations, systems can track sensitivity and apply noise mechanisms with formal guarantees. Frameworks such as OpenDP \cite{opendp_opendp_2020} and Tumult Analytics \cite{berghel_tumult_2022} have demonstrated that this ``correct-by-construction'' approach can scale to multiprocess and multihost environments. However, the primary limitation of this paradigm is its integration into existing environments: integrating these DSLs into pre-existing, large-scale data pipelines often requires a significant rewrite of the underlying semantics. Consequently, many real-world systems continue to rely on ad-hoc or custom implementations that lack these safety guarantees, making the need for easy-to-use testing frameworks such as ours essential.

\paragraph{Formal Verification.} A significant body of research in DP has focused on formal languages and static analysis \cite{gaboardi2013linear,10.1145/3009837.3009884,10.1145/3360598,10.1007/978-3-031-30044-8_17}. These systems employ programming language techniques to statically verify that an algorithm formally satisfies the mathematical definitions of DP. The primary benefit of this approach is providing a provable guarantee of privacy, effectively eliminating entire classes of bugs at compile-time. Despite these robust guarantees, such formal languages have seen little practical adoption, primarily due to significant usability and integration barriers. These systems often impose a steep learning curve, requiring practitioners to become specialized in non-mainstream programming languages that are difficult to integrate into existing, mature data science pipelines.

\section{Framework}
\label{sec:motivation}

In this section, we introduce our gray-box auditing framework, \methodtrusted~and \methoduntrusted. We begin by outlining the key insights that make our fast, \emph{structural} verification possible.

\subsection{Motivation \& Insights}
The transition of DP from theoretical research to production software has revealed a significant gap in our testing capabilities. While the mathematical properties of standard privacy primitives (e.g., the Laplace and Gaussian mechanisms) are well-understood, the pipelines that utilize them are increasingly complex and prone to implementation errors.

In practice, most privacy violations do not stem from flaws in the primitives themselves. Rather, they arise from mechanism-independent integration errors, including sensitivity underestimation, incorrect noise scaling, and improper accounting. These are logic bugs, not statistical ones. Still, the standard approach to detecting them remains distributional auditing, which treats the algorithm as a black box and attempts to distinguish its output distributions on neighboring datasets. This approach faces three challenges: 
\begin{enumerate} 
    \item \textbf{Statistical Intractability:} As noted by \citet{gilbert2018property}, auditing unconstrained algorithms is computationally intractable. In complex pipelines involving adaptive composition, the output space grows exponentially.
    
    \item \textbf{Complexity of MIA Design:} To detect a violation, an auditor must design a Membership Inference Attack (MIA) capable of distinguishing distributions with high confidence. This requires specialized expertise in threat modeling, extensive hyperparameter tuning, and massive computational resources.
    \item \textbf{Lack of Error Source:} Even if the MIA successfully detects a violation, it provides a binary fail signal. It cannot pinpoint whether the error lies in pre-processing, sensitivity calibration, privacy primitives, or post-processing.
\end{enumerate}

\paragraph{A Structural Decomposition Approach} To address this, we move beyond black-box distributional auditing and instead leverage the program's structure to detect integration errors deterministically. Our framework initially assumes that the privacy primitives are correctly implemented—an assumption we later relax. This is a reasonable premise given the availability of community-vetted, open-source libraries such as OpenDP \cite{opendp_opendp_2020} and Tumult Analytics \cite{berghel_tumult_2022}. Then, if a mechanism such as \texttt{laplace\_mechanism(input, sensitivity, scale)} is correctly implemented, we can focus on how the mechanism is \emph{used} within the broader pipeline.

Observing that DP algorithms interleave private primitives with logic that is not supposed to depend on the private data, we approach the auditing problem through \emph{structural decomposition}. We distinguish between data-dependent functions (the privacy primitives) and data-independent ones (the control flow, pre-processing, and post-processing logic). Data-dependent functions are the only legitimate source of variation between executions on $D$ and $D'$, and all other variations must be explained through sensitivity. To verify this, consider a thought experiment: suppose we overwrite every data-dependent mechanism in a pipeline to return a constant value (frozen output). If we then execute this pipeline on $D$ and $D'$ with the same seed, the program's state must remain identical throughout execution. We define this property as \emph{invariance under $D$ and $D'$}. If any intermediate value diverges, the function was not data-independent, implying a privacy leak in the processing logic.

This decomposition transforms the auditing problem. Instead of designing MIAs to estimate privacy loss bounds, we need to answer the following questions: (i) Do the data-independent parts of the pipeline remain \emph{invariant} when the dataset changes, given fixed mechanism outputs? and (ii) Do the inputs fed into the trusted mechanisms respect the desired overall privacy guarantees?

This shift in perspective enables us to transition from computationally expensive attacks to fast unit tests suitable for standard software development pipelines. This logic forms the foundation of \methodtrusted, which we detail in the following section.

\subsection{\methodtrusted}
\label{sec:lightweight-auditing}

Before presenting our method, we first outline the core assumptions and the threat model under which our auditing framework operates.

\begin{assumption}
    \label{assumption:correct-mechanisms}
    We assume the availability of correctly implemented privacy primitives (e.g., Gaussian, Laplace, Exponential mechanisms) for which the privacy accounting rules under composition are well established, and we have access to a correct accountant for them. 
\end{assumption}

\begin{assumption}
    \label{assumption:randomness-controlability}
    We assume access to a function that snapshots and restores the state of the system's pseudo-random number generator. We assume that the developers provide this function, as it is becoming increasingly standard for reproducibility purposes. 
\end{assumption}

\paragraph{Threat Model: Gray-Box Auditing.} We adopt a gray-box auditing model in which the adversary is not modeled as a malicious external actor, but rather as a developer who inadvertently introduces implementation errors into the processing logic. We designate this approach as ``gray-box'' because it relies on a selective visibility of the codebase: we require explicit access to the algorithm's control flow, the intermediate variables passed to privacy primitives, and the state of the pseudo-random number generator. However, we treat the internal implementations of the privacy primitives themselves (e.g., the specific logic within a Laplace mechanism) as a black-box. This distinction is crucial, as it allows us to leverage structural knowledge of the intermediate code.

\paragraph{Intuition}
Our approach operates on the principle that in a correctly implemented DP algorithm, the dependency on private data should be restricted to specific, accounted-for primitives. All surrounding control flow and parameter logic must remain data independent. To verify this, we employ a strategy of \emph{isolation}: if we freeze the outputs of the privacy primitives, the entire remaining execution path of the algorithm should be identical for any two neighboring datasets $D$ and $D'$. Any deviation in the program state outside of these primitives signals that private data has leaked into the ``public'' logic without paying the necessary privacy budget. This reduces the complex auditing task to two straightforward verifications: (i) that the data-independent logic remains \emph{invariant}, and (ii) that the inputs entering the privacy primitives adhere to their declared sensitivity limits.

\paragraph{Trace Generation and Strict Equality.}
To implement this verification, \methodtrusted{} (formalized in \Cref{alg:record-play}) generates parallel execution traces by instrumenting the target algorithm $\mathcal{A}$. A central hook intercepts all calls to DP primitives. 
Beyond standard primitives (e.g., Laplace), our framework supports a specialized verification node denoted as \texttt{ensure\_equality}. This node allows the auditor to explicitly assert that specific intermediate values, such as hyperparameters, metadata-derived constants, or loop bounds, remain equal across runs. This ensures that changes to private data do not affect data-independent components of the computation. The framework operates in two distinct phases:

\paragraph{Phase 1: Recording on Initial Dataset ($D$)}
First, we execute the instrumented algorithm $\mathcal{A}$ on an initial dataset, $D$. The instrumentation hook (\Cref{alg:record-play}, line 10) operates in \texttt{RECORD} mode. For each invocation of a DP primitive $\mathcal{M}$ (or equality check), the hook logs a detailed entry into a trace $\mathcal{T}$, containing: (1) the mechanism type, (2) its parameters $\rho$ (e.g., noise variance $\sigma$), (3) the input query data $q$ (specifically $q_D$), (4) the PRNG state, and (5) the exact output $o$ produced. This generates an ordered log of every stochastic decision point and equality assertion within the execution path.

\paragraph{Phase 2: Replaying on Neighboring Dataset ($D'$)}
Next, we execute $\mathcal{A}$ on a neighboring dataset, $D'$ (lines 5-7), with the hook in \texttt{REPLAY} mode. When a primitive is called, it performs the following logic:

\begin{enumerate}
    \item \textbf{Verifies Control-Flow:} It checks that the current call type and order match the corresponding call in the recorded trace $\mathcal{T}$. A mismatch flags a violation, indicating the control flow is data-dependent.
    \item \textbf{Verifies Invariance:} If the node is of type \texttt{ensure\_equality} (lines 27-30), it asserts that the input $q$ matches the recorded input from the $D$ trace. If they differ, it flags a violation. For privacy primitives, invariance is checked on non-sensitive arguments (e.g., declared noise scale).
    \item \textbf{Logs Inputs:} It logs the new inputs $q_{D'}$ into a new trace $\mathcal{T}'$.
    \item \textbf{Enforces Determinism:} It restores the PRNG state to the \textit{post-execution} state captured during the recording phase and returns the \emph{frozen} output from the $D$ trace.
\end{enumerate}

By forcing the mechanisms to return the same output under both $D$ and $D'$, we isolate the valid source of variation. Finally, the auditor compares the inputs $q_D$ and $q_{D'}$ stored in the traces. If the empirical distance between them exceeds the declared sensitivity (and yet the accounting claimed otherwise), a sensitivity violation is flagged.

\begin{algorithm}
\caption{Trace Generation via \methodtrusted}
\label{alg:record-play}
\begin{algorithmic}[1]
\Statex \textbf{Input:} Instrumented Algorithm $\mathcal{A}$, neighboring datasets $D, D'$
\Statex \textbf{Output:} Execution traces $\mathcal{T}$, $\mathcal{T}'$

\Procedure{GenerateTraces}{$\mathcal{A}, D, D'$}
    \Statex \emph{// Phase 1: Record}
    \State $\mathcal{T} \leftarrow []$ \Comment{Initialize trace for $D$}
    \State \textsc{SetInstrumentationMode}(\texttt{RECORD}, $\mathcal{T}$)
    \State $\mathcal{A}(D)$ \Comment{Run on $D$; hook populates $\mathcal{T}$}

    \Statex \emph{// Phase 2: Replay}
    \State $\mathcal{T}' \leftarrow []$ \Comment{Initialize trace for $D'$}
    \State \textsc{SetInstrumentationMode}(\texttt{REPLAY}, $\mathcal{T}$, $\mathcal{T}'$)
    \State $\mathcal{A}(D')$ \Comment{Run on $D'$; hook populates $\mathcal{T}'$}
    
    \State \textbf{return} $\mathcal{T}, \mathcal{T}'$
\EndProcedure

\Statex
\Procedure{OnPrimitiveCall}{$\mathcal{M}, \rho, q$} \Comment{Instrumentation Hook}
    \If{\texttt{mode} = \texttt{RECORD}}
        \If{$\mathcal{M} = \texttt{ensure\_equality}$}
             \State $o \leftarrow q$ \Comment{Pass-through for equality check}
        \Else
             \State $o \leftarrow \mathcal{M}(q, \rho)$ \Comment{Execute real mechanism}
        \EndIf
        \State $s_{post} \leftarrow$ \textsc{GetRNGState}()
        \State $\mathcal{T}$.\textsc{add}($(\mathcal{M}, \rho, q, s_{post}, o)$) \Comment{Log full state}
        \State \textbf{return} $o$
    \ElsIf{\texttt{mode} = \texttt{REPLAY}}
        \State $k \leftarrow |\mathcal{T}'| + 1$ \Comment{Current call index}
        \If{$k > |\mathcal{T}|$ \textbf{or} $\mathcal{M} \neq \mathcal{T}[k].\mathcal{M}$}
            \State \Comment{Control flow mismatch; traces are now invalid.}
            \State \textbf{return} \texttt{STOP} 
        \EndIf
        
        \State $(\dots, s_{post}, o_{rec}) \leftarrow \mathcal{T}[k]$ \Comment{Read from $D$ trace}

        \If{$\mathcal{M} = \texttt{ensure\_equality}$}
            \If{$q \neq \mathcal{T}[k].q$} \Comment{Verify Invariance}
                \State \textbf{return} \texttt{STOP} 
            \EndIf
            \State $\mathcal{T}'$.\textsc{add}($(\mathcal{M}, q)$) \Comment{Log $D'$ inputs/params}
            \State \textbf{return} $q$
        \EndIf
        
        \State $\mathcal{T}'$.\textsc{add}($(\mathcal{M}, \rho, q)$) \Comment{Log $D'$ inputs/params}
        \State \textsc{SetRNGState}($s_{post}$) \Comment{Restore post-call RNG state}
        \State \textbf{return} $o_{rec}$ \Comment{Return \emph{frozen} output from $D$ run}
    \EndIf
\EndProcedure
\end{algorithmic}
\end{algorithm}

In the next section, we generalize \methodtrusted to scenarios where we do not trust the underlying implementation of the privacy primitives, enabling us to also find bugs in their accounting.

\subsection{Distributional Audit: \methoduntrusted}
\label{sec:distributional-audit}

The lightweight audit in \Cref{sec:lightweight-auditing} is a testing tool for verifying consistency and data-independence. However, it relies on \Cref{assumption:correct-mechanisms}: the privacy primitives ($\mathcal{M}_i$) themselves are correctly implemented and their accounting is accurate.

In this section, we lift this assumption. We consider a scenario in which the privacy primitives themselves are untrusted because they have a custom, unvetted implementation, are part of a complex legacy system, or use non-standard theoretical privacy accounting.

To handle this, we introduce \textbf{\methoduntrusted}, extending our lightweight framework into a \emph{gray-box distributional audit}.
Unlike black-box approaches that must treat the whole algorithm as opaque, we leverage the \methodtrusted\ trace to isolate each mechanism and its exact sensitive inputs, and then perform a statistical audit on each isolated component. The core idea is to move from verifying \emph{declared} properties to empirically \emph{estimating} the privacy loss. Our method (formalized in \Cref{alg:record-play-sample}) works in three phases:

\begin{enumerate}
    \item \textbf{Phase 1: Record \& Replay.} See ~\Cref{sec:lightweight-auditing}.

    \item \textbf{Phase 2: Sample.} With the inputs for each primitive isolated, we now treat each \emph{individual} $\mathcal{M}_i$ as a black box. For each tuple $(\mathcal{M}_i, q_{D,i}, q_{D',i})$ in the trace $\mathcal{T}$, we perform a statistical sampling experiment. We execute $\mathcal{M}_i(q_{D,i})$ $N$ times to generate an empirical output distribution $\hat{P}_i$, and similarly execute $\mathcal{M}_i(q_{D',i})$ $N$ times to generate $\hat{P}'_i$.

    \item \textbf{Phase 3: Estimate \& Compose.}
    At this stage, auditing the isolated component $\mathcal{M}_i$ reduces to a standard black-box auditing problem. As a result, our framework can incorporate any existing distributional auditing technique \cite{jagielski2020auditingdifferentiallyprivatemachine, nasr_adversary_2021, nasr2023tight}. Building on this flexibility, we propose a specific auditing technique based on Privacy Loss Distributions (PLDs). From each pair of empirical distributions $(\hat{P}_i, \hat{P}'_i)$, we compute an empirical PLD $\mathcal{L}_i $ following the methodology of~\cite{doroshenko2022connect}.
    Given the set of empirical PLDs ${\mathcal{L}_1, \dots, \mathcal{L}_k}$ for the entire execution path, we then use a numerical privacy accountant \cite{doroshenko2022connect} to compose them and obtain a single end-to-end $(\varepsilon, \delta)$ bound for the full algorithm $\mathcal{A}$ on this specific path and input pair. Additional details on the PLD estimation procedure are provided in \Cref{appendix:pld_estimation}.
\end{enumerate}

This method provides a concrete, empirical estimator of the privacy loss, effectively bypassing the reliance on the algorithm's internal accounting or implementation correctness.

This approach differs from standard black-box distributional auditing, which typically requires re-executing the entire pipeline to generate inputs for each sample. Instead, we leverage the trace to fix inputs for specific privacy primitives, decoupling expensive preprocessing from the audit. By isolating each primitive, we avoid the computational cost of the full end-to-end function and focus on decomposed, cheap-to-execute privacy primitives. This efficiency is critical to our goal of providing a tool suitable for integration into CI/CD pipelines, where audits must be frequent and low-latency.

Interestingly, this framework supports hybrid composition. Because we introduce a novel way of estimating a PLD for an arbitrary mechanism without assuming a priori structure (see ~\Cref{appendix:pld_estimation}), we can interleave \emph{trusted} primitives with \emph{untrusted} ones (audited empirically). This flexibility allows practitioners to allocate their computational auditing budget where it is needed most, minimizing overall testing costs while maximizing coverage.

\begin{algorithm}
\caption{Distributional Audit via \methoduntrusted}
\label{alg:record-play-sample}
\begin{algorithmic}[1]
\Statex \textbf{Input:} Algorithm $\mathcal{A}$, neighboring datasets $D, D'$, Sample count $N$ (for statistical audit), expected $(\varepsilon, \delta)$
\Statex \textbf{Output:} A bolean answer outputting if the empirical privacy guarantee $\hat{\varepsilon}$ satisfies the claimed $\varepsilon$ at the same $\delta$

\Procedure{AuditDistributional}{$\mathcal{A}, D, D', N$}
    \State \emph{ Phase 1: Record \& Replay (from \Cref{sec:lightweight-auditing})}
    \State $\mathcal{T}, \mathcal{T'} \leftarrow$ \textsc{RunRecordReplay}($\mathcal{A}, D, D'$)
    \If{$\mathcal{T}$ contains control-flow mismatch}
        \State \textbf{return} \texttt{STOP}
    \EndIf

    \State $\mathcal{L}_{list} \leftarrow []$ \Comment{List to store empirical PLDs}

    \Statex \emph{// Phase 2: Sample}
    \For{each mechanism call $i=1 \dots |\mathcal{T}|$}
        \State $(\mathcal{M}_i, q_{D,i}, q_{D',i}) \leftarrow \mathcal{T}[i]$
        
        \State $O_{D}, O_{D'} \leftarrow [], []$ \Comment{Outputs from $D$ and $D'$}
        \For{$j=1 \dots N$} \Comment{Collect $N$ samples}
            \State $O_{D}$.\textsc{add}($\mathcal{M}_i(q_{D,i})$) \Comment{Run on $D$ input}
            \State $O_{D'}$.\textsc{add}($\mathcal{M}_i(q_{D',i})$) \Comment{Run on $D'$ input}
        \EndFor
        
        \Statex \emph{// Phase 3: Estimate}
        \State $\hat{P}_i, \hat{P}'_i \leftarrow$ \textsc{BuildEmpiricalDistributions}($O_{D}, O_{D'}$)
        \State $\mathcal{L}_i \leftarrow$ \textsc{EstimatePLD}($\hat{P}_i, \hat{P}'_i$) \Comment{e.g.,\cite{jagielski2020auditingdifferentiallyprivatemachine}}
        \State $\mathcal{L}_{list}$.\textsc{add}($\mathcal{L}_i$)
    \EndFor

    \Statex \emph{// Phase 3: Compose}
    \State $\mathcal{L}_{total} \leftarrow$ \textsc{NumericalAccountant}($\mathcal{L}_{list}$) \Comment{e.g., via \cite{doroshenko2022connect}}
    \State $\hat{\varepsilon} \leftarrow \textsc{GetPrivacyBound}(\mathcal{L}_{total}, \delta)$
    \State \textbf{return} $\hat{\varepsilon} < \varepsilon$
\EndProcedure
\end{algorithmic}
\end{algorithm}

\section{Evaluation}
\label{sec:evaluation}

To validate the practicality of our framework, we designed an evaluation benchmark to highlight that our proposed approach is: (i) \emph{Effective:} Our framework detects real, subtle, and diverse classes of DP implementation bugs; and (ii) \emph{Efficient:} Our framework is computationally lightweight. First, we detail the implementation of our auditing tool and our strategy for generating the neighboring datasets $(D,D')$ used to trigger privacy violations. Then, we discuss our findings applying both \methodtrusted ~ and \methoduntrusted.

\subsection{Implementation}

We implemented our framework as a Python library that integrates with the \texttt{pytest} testing ecosystem. This design choice positions privacy auditing as a standard component of the software development lifecycle, capable of running alongside unit and integration tests. The core instrumentation logic (the \texttt{OnPrimitiveCall} hook from \Cref{alg:record-play}) uses context variables \cite{noauthor_contextvars_nodate}. Crucially, our framework introduces no side effects or performance overhead in production; tracing is activated solely within the testing environment.

\Cref{fig:simple_example} shows a fully self-contained integration example. A Python decorator marks the DP primitives, in this case, just the Laplace Mechanism (line 1). A simple \texttt{pytest} test (line 24) instantiates two neighboring datasets $ D$ and $ D'$, as well as the auditor, which is in $\texttt{record}$ mode by default. The private algorithm $\texttt{private\_function}$ (line 11) is executed twice---once on $D$ and once on $D'$---with the second execution in $\texttt{replay}$ mode (line 32). We use $\texttt{ensure\_equality}$ (line 13) to check for data-dependent changes in parameters. Afterwards, the auditor checks for mismatches in the execution trace in $\texttt{validate\_records}$ (line 37) and then starts the $\texttt{distributional\_audit}$. Finally, the test concludes by deriving an overall privacy bound (see \autoref{appendix:pld_estimation}) and asserting whether it exceeds the declared privacy budget. While Re:cord-play (\Cref{sec:lightweight-auditing}) fixes the mechanism's output during replay, it also checks inputs: it calculates the empirical distance between inputs across the two runs and compares this against the declared sensitivity in $\texttt{validate\_records}$. In the example, the input $\texttt{scaled\_count}$ differs by 2 between runs. Our framework catches the bug and pinpoints the violation. In this case, it detects that there is a sensitivity miscalibration when the $\texttt{LM}$-mechanism if called for the first time.

The example illustrates the framework's ease of integration into standard software development workflows. The integration consists of two components: (i) writing tests and (ii) marking privacy primitives in the library. For part (i), shown in \Cref{fig:simple_example}, the auditing logic requires minimal boilerplate code: the developer simply wraps the target algorithm within the Auditor scope, first in \texttt{RECORD} mode to capture the execution trace $\mathcal{T}$ on $D$, and subsequently in \texttt{REPLAY} mode to generate $\mathcal{T}'$ on $D'$. The API automatically handles internal state tracking and side-effect isolation, allowing practitioners to seamlessly perform deterministic sensitivity checks (\texttt{validate\_records}) and statistical audits (\texttt{distributional\_\allowbreak audit}) within existing testing pipelines.

Integrating our library is designed to be non-intrusive. For part (ii), developers mark privacy primitives using our \texttt{@audit\_spec} decorator and check for invariance using the \texttt{ensure\_equality} functions, as illustrated in \Cref{fig:simple_example}. The decorator specifies a mapping of the arguments required to check privacy violations. The \texttt{audit\_spec} takes the following arguments: (i) \texttt{kind}, which sets the mechanism's name for debugging and reporting; (ii) \texttt{input\_arg}, denoting the data that the mechanism applies noise to; (iii) \texttt{sensitivity\_arg}, denoting an upper bound on the allowed change in \texttt{input\_arg}; and (iv) \texttt{metric\_fn}, which defines the distance metric used to compute sensitivity across two runs on neighboring datasets. Optionally, developers may provide (v) a privacy \texttt{accountant}. When an accountant is provided, the primitive is treated as trusted, and our tool verifies it using \methodtrusted. Otherwise, as is common for custom mechanisms, the primitive defaults to \methoduntrusted, triggering statistical verification. This annotation process is performed only \textit{once} per library. In our experiments, we utilize the numerical accountant from \citet{doroshenko2022connect}.

\begin{figure}
\caption{End-to-end testing example of a code with a sensitivity miscalibration.}
\label{fig:simple_example}
\begin{lstlisting}[
    style=mypython,       % Use the Python style we defined
    escapechar=€,         % Set '€' as the escape character
    captionpos=b,          % Ensure caption is at the bottom
    xleftmargin=1.5em,
    framexleftmargin=1.5em,
]
@audit_spec(
    kind="LM",
    input_arg="x",
    sensitivity_arg="sensitivity",
    metric_fn=dist_l1,
    accountant=laplacian_pl
)
def laplace_mechanism(x, sensitivity, epsilon):
    return x + sample_laplacian_noise(scale=sensitivity / epsilon)

def private_function(data, multiplier, epsilon):
    # ensure multiplier is not data-dependent
    ensure_equality(m = multiplier)

    # sensitivity = 1 (under add/remove)
    count = len(data)
    # sensitivity = multiplier
    scaled_count = count * multiplier

    # BUG: incorrect sensitivity used
    noisy_count = laplace_mechanism(scaled_count, sensitivity=1, epsilon=epsilon)
    return noisy_count
    
def test_simple_audit():
    D = [0,0,0]     # D dataset
    Dp = [0,0,0,0]  # D' dataset
    eps = 1.0
    auditor = Auditor()
    with auditor:
        _ = private_function(D, 2, eps)
        
    auditor.set_replay()
    with auditor:
        _ = private_function(Dp, 2, eps)

    # Run Re:cord-play
    auditor.validate_records()

    # Run Re:cord-play-sample
    eps_rec= auditor.distributional_audit(delta=1e-6, n_samples=1e5)
    assert eps_rec <= eps, "Privacy violation"
\end{lstlisting}
\end{figure}

\subsection{Datasets}
In this section, we describe the data used to run our auditing. We want to stress that the goal is to keep the execution time and memory usage as small as possible.

\paragraph{Synthetic Data} To create a controlled and reproducible test environment, our synthetic data procedure generates an initial dataset $D$ as a low-dimensional tabular dataset. Using a seeded pseudo-random number generator, it creates a table with a fixed number of records (e.g., $n=200$) and a small number of categorical attributes (e.g., 2 columns). The values for these attributes are drawn uniformly at random from small, predefined integer ranges.

\paragraph{Neighboring Datasets} A core component of any DP audit is to generate neighboring datasets, as the definition of DP is parameterized by the relationship between $D$ and $D'$. Our evaluation must test an implementation's robustness against the specific adjacency model (i.e., the definition of "neighbor") it claims to support. DP literature primarily considers two types of dataset adjacency, corresponding to different privacy models and threat scenarios:
\begin{enumerate}
    \item \textbf{Unbounded Adjacency (Add/Remove):} The most common definition, used when the dataset size $n$ is not public. Two datasets $D, D'$ are neighbors if one can be formed from the other by adding or removing a record. This models the privacy risk to an individual's data based on whether they participate in the analysis or not. The sensitivity of a query (e.g., SUM) under this model is typically its maximum possible contribution (e.g., the clipping bound $C$).
    \item \textbf{Bounded Adjacency (Replace-One):} This definition is standard where dataset size n is public. Two datasets $D, D'$ are neighbors if they both have exactly $n$ records, and $D'$ can be formed by replacing one record in $D$ with any other possible record from the data universe. This models the privacy risk to an individual when they change their data. The sensitivity of a query under this model can be different (e.g., up to 2C for SUM).
\end{enumerate}

\paragraph{Adjacency Agnosticism} A key advantage of our framework is its independence from specific adjacency definitions. Although the previous two notions are the most common, alternative notions exist, notably in graph data or machine learning \cite{pmlr-v139-kairouz21b}. Our tool enforces no specific model. Instead, it evaluates consistency against any provided neighboring pair $(D,D')$.

\paragraph{Generation Strategies} To instantiate the necessary dataset pairs $(D,D')$ for auditing, we employ a diverse set of heuristics ranging from standard mutations to adversarial edge cases. For unbounded adjacency, we randomly select candidates, while additions are generated via diverse strategies: uniform sampling, marginal sampling, or data duplication. To stress-test numerical stability and overflow handling, we inject values near floating-point limits (e.g., \texttt{max(float64)}). Furthermore, we explicitly target domain inference vulnerabilities by sampling values outside the implied min/max range of $D$.
For bounded adjacency, we combine the add and remove operations to form neighbor pairs. As discussed in \Cref{sec:futurework}, dataset generation is outside the scope of this work; our framework is designed to integrate seamlessly with recent advances in this area (e.g., \cite{9519405}).

\subsection{Auditing Results}
\label{sec:auditing-results}

Before presenting our results, we clarify two points: (i) \emph{Intent}: Our goal is to motivate the adoption of our automated gray-box testing framework, not to criticize library maintainers. The subtleties of DP logic make manual verification challenging, even for experienced teams. (ii) \emph{Completeness}: For each audited implementation, we terminate the audit at the first detected bug. Consequently, our results are not exhaustive; instead, they are meant to illustrate the practical applicability of our framework with minimal engineering effort, rather than to catalog all potential issues.

\paragraph{Reproducibility} We ensure reproducibility by reporting the specific Git commit hash for every codebase audited and associated links in~\Cref{tab:libraries} of the Appendix. This list serves as the ground truth for our findings, allowing independent verifiers to inspect the exact logic present at the time of the audit. 

We selected diverse target implementations from widely used, open-source DP libraries. Our results are summarized in \Cref{tab:dp_bugs}, with details in \Cref{appendix:audited_codebases}.
We identified privacy violations ranging from subtle preprocessing flaws to misapplication of parameters. Below, we detail specific bugs and their implications.

\subsubsection{SmartNoise SDK}

We audited the SmartNoise SDK's private covariance matrix implementation, a key component of algorithms such as Logistic Regression. We identified a bug  (via \methodtrusted) where the covariance computation incorrectly uses the \emph{unprocessed} data, even after a sanitization step was intended to be applied. As shown in \Cref{fig:smartnoise-bug}, the function creates \texttt{newdata} by sanitization but then computes \texttt{trueval} using the original \texttt{data}. This is a \emph{sensitivity miscalibration}: the declared sensitivity \texttt{self.sens} is calculated assuming censored data, but an uncensored outlier can cause the true sensitivity to be arbitrarily large. Our framework detects this in a case where a neighboring dataset $D'$ contains large outliers. The \methodtrusted{} hook logs the data-dependent inputs to the noise primitive, $q_D$ and $q_{D'}$. In the validation phase, the framework computes the empirical distance $\| q_D-q_{D'}\|_1$ and finds it is greater than the \emph{declared} sensitivity \texttt{self.sens}, returning an error. 

\begin{figure}
\caption{Simplified code for the Sensitivity Bug in the SmartNoise SDK Covariance estimation. The declared sensitivity \texttt{self.sens} is calculated assuming censored data, but the \texttt{covar} function receives the original, uncensored \texttt{data}.} 
\label{fig:smartnoise-bug}
\begin{lstlisting}[
    style=mypython,       % Use the Python style we defined
    escapechar=@,         % Set '@' as the escape character
    captionpos=b          % Ensure caption is at the bottom
]
def release(self, data):
    newdata = censor_data(data)
    newdata = fill_missing(new_data)
    ... 
    @\colorbox{red!20}{    BUG: 'data' is used instead of 'newdata'}@
    @\colorbox{red!20}{    trueval = covar(data.values.T, self.intercept)}@
    scale = self.sens / self.epsilon
    val = np.array(true_val) + dp_noise(n=len(true_val), noise_scale=scale)
    return list(val)
\end{lstlisting}
\end{figure}

\subsubsection{Smartnoise SQL} We audited the SmartNoise SQL framework and uncovered a miscalculation in the privacy accountant via \methoduntrusted. As shown in Figure~\ref{fig:smartnoise-odometer-bug}, the implementation of the homogeneous privacy odometer omits the
$\log(1/\delta)$ term required by the advanced composition bound of \cite{dwork2010boosting,kairouz2015composition}, replacing it with a linear $1/\delta$ factor. This error dramatically inflates the ``optimal'' composition term $b$, causing it to grow by orders of magnitude for standard parameter ranges. As a result, the intended advanced composition bound is never applied: the accountant always falls back to more conservative alternatives, yielding substantially looser privacy guarantees than theoretically possible. This example highlights a notable capability of our framework: it can be used to detect implementation bugs that are not necessarily privacy violations. By comparing our estimated privacy loss with the accountant's reported upper bound, a significant discrepancy signals a potential bug, even if no privacy violation occurs.

\begin{figure}
\caption{SmartNoise SQL missing $\log(1/\delta)$ factor in the
homogeneous odometer.}
\label{fig:smartnoise-odometer-bug}
\begin{lstlisting}[
    style=mypython,       % Use the Python style we defined
    escapechar=@,         % Set '@' as the escape character
    captionpos=b          % Ensure caption is at the bottom
]
class Odometer:
    def spent(self):
        ...
        @\colorbox{red!20}{BUG: equation missing log term around (1/tol)}@
        @\colorbox{red!20}{optimal\_b = optimal\_left\_side + epsilon * np.sqrt(}@
                    @\colorbox{red!20}{2 * self.k * (1/tol)}}
\end{lstlisting}
\end{figure}

\subsubsection{Synthcity}

We audited the PrivBayes implementation \cite{10.1145/3134428} and found two distinct bugs using \methodtrusted. First, as shown in \Cref{fig:synthcity-bug1}, the code uses the output of the Exponential Mechanism (\texttt{candidate\_idx}) to index a \emph{private}, un-noised list (\texttt{mutual\_info\_\allowbreak list}), and then uses this private value in a public \texttt{if} statement. This constitutes an \emph{Invariant Violation}, as the execution path (whether the \texttt{if} block is entered) leaks information. Our framework catches this by setting an \texttt{ensure\_equality} node on the right-hand term of the \texttt{if} condition. During \texttt{REPLAY} on $D'$, the hook (\Cref{alg:record-play}) detects that the sequence of mechanism calls differs from the trace on $D$. Second, in the utility function (\texttt{\_laplace\_noise\_parameter} in \Cref{fig:synthcity-bug1}), we found a \emph{Noise Miscalibration} bug. The Laplace noise scale numerator can become zero if \texttt{self.K == n\_features}. This disables noise addition entirely, releasing the true value. This bug is detected by both \methodtrusted~ and \methoduntrusted, which report an $\varepsilon$ value larger than the specified one. We also audited their AIM \cite{mckenna_aim_2024} implementation and found no issues to report.

\begin{figure}
\caption{Simplified code for the two bugs in Synthcity's DP Bayes algorithm. First, the output of the Exponential Mechanism, \texttt{candidate\_idx}, is used to index into a private list, and the result is used in a non-private \texttt{if} check. Also, \texttt{self.K} (e.g., the number of parents) equals \texttt{n\_features}, and the noise scale becomes zero, completely disabling privacy protection.}
\label{fig:synthcity-bug1}
\begin{lstlisting}[
    style=mypython,       % Use the Python style we defined
    escapechar=@,         % Set '@' as the escape character
    captionpos=b          % Ensure caption is at the bottom
]
def _greedy_bayes(self, data):
        ...
        sampling_distribution = self._exponential_mechanism(
            data,
            parents_pair,
            mutual_info)
        candidate_idx = np.random.choice(
            list(range(len(mutual_info))), 
            p=sampling_distribution)
        sampled_pair = parents_pair[candidate_idx]
        @\colorbox{red!20}{    BUG: indexed in private data}@
        @\colorbox{red!20}{    if self.mi\_thresh >= mutual\_info[candidate\_idx]:}@
            sampled_pair = network_edge(sampled_pair.feature, parents=[])
        ...

def _laplace_noise_parameter(self, n_items: int, n_features: int) -> float:
    @\colorbox{red!20}{BUG: self.K can be equal to n\_features}@
    @\colorbox{red!20}{return 2*(n\_features - self.K) / (n\_items * self.epsilon)}@
\end{lstlisting}
\end{figure}

\subsubsection{Diffprivlib}
We audited the \texttt{LinearRegression} model, which fits an ordinary least-squares linear regression to training data in a differentially private manner. We found a \emph{Sensitivity Miscalibration} via \methodtrusted~ in the \texttt{\_construct\_regression\_obj} function. For the diagonal values of the quadratic term in the squared error, i.e. \texttt{mono\_coef\_2[i, i]}, the sensitivity is the squared maximum of the lower and upper bound of the feature. However, as shown in \Cref{fig:diffprivlib-sens-bug}, only the lower bound is being used.  Our framework caught this by checking whether the empirical sensitivity exceeds the recorded sensitivity.

\begin{figure}
\caption{Simplified code of the Sensitivity Miscalibration bug in Diffprivlib's \texttt{\_construct\_regression\_obj} function. The computation might incorrectly result in zero and is passed into the Laplace mechanism.}
\label{fig:diffprivlib-sens-bug}
\begin{lstlisting}[
    style=mypython,       % Use the Python style we defined
    escapechar=@,         % Set '@' as the escape character
    captionpos=b          % Ensure caption is at the bottom
]
def _construct_regression_obj(X, y, bounds_X, bounds_y, epsilon, alpha, random_state):
    ...
    for i in range(n_features):
        @\colorbox{red!20}{ BUG: lower bound is used twice}@
        @\colorbox{red!20}{ sensitivity = np.max(np.abs([bounds\_X[0][i],}@
                    @\colorbox{red!20}{bounds\_X[0][i]])) ** 2}@
        mech = LaplaceFolded(epsilon=local_epsilon, sensitivity=sensitivity, lower=0, upper=float("inf"), random_state=random_state)
        mono_coef_2[i, i] = mech.randomise(coefs[2][i, i])
    ...
\end{lstlisting}
\end{figure}

We also uncovered an invariance violation in \texttt{diffprivlib}'s \texttt{LogisticRegression} classifier via \methodtrusted. Internally, the implementation infers the label set via \texttt{classes = np.unique(y)}, so under the add/remove adjacency model, the presence or absence of a single sample from a rare class can change \texttt{classes} between neighboring datasets. In our \methodtrusted{} audit, this causes a control-flow divergence and violates our invariance check. Conceptually, this leaks whether a given label appears at least once in the training data (label-space domain inference). Instead, the set of possible classes should be treated as public and set by the user.

We found no issues to report in the \texttt{histogram} tool, \texttt{GaussianNB}, \texttt{StandardScaler}, \texttt{PCA}, and \texttt{KMeans}  models.

\subsubsection{Private-PGM}
We audited the JAM \cite{fuentes2024jamcode} algorithm, which utilizes the bounded DP model. Under this definition, changing a single record shifts a marginal histogram by at most 2 in $\ell_{1}$-distance. JAM selects marginals via the Exponential Mechanism, using the scoring function:
\begin{equation*}
    \textstyle \texttt{model\_error}[m] \;-\; \texttt{pub\_errors}[m]
\end{equation*}
Both terms depend on the private inputs and can shift by up to 2. Because these changes can occur in opposite directions, the true sensitivity is 4 (i.e., $\textstyle |\Delta (\texttt{model\_error}[m] - \texttt{pub\_errors}[m])| \le 4$). The implementation, however, underestimates this value, leading to a sensitivity misscalibration bug detected via \methodtrusted. This appears to be a regression introduced during the migration to the Private-PGM package, as the original research code correctly sets \texttt{score\_sensitivity = 4}, see \citet{fuentes2024jamcode}. We have also audited their MWEM+PGM, Gaussian+AppGM, MST, and AIM algorithms and found no issues.

\subsubsection{MOSTLY AI Engine} Our audit of the MOSTLY AI synthetic data engine identified two classes of issues: incorrect budget composition (detected via \methoduntrusted) and data-dependent control flow (detected via \methodtrusted). The numeric encoder applies two separate DP mechanisms to the same column: one for approximating bounds and one for identifying non-rare values, each spending the same amount of budget. Only a single budget is accounted for, although both routines add Laplace noise and therefore each consumes $\varepsilon$. In cases where both branches are activated (e.g., low-cardinality numeric columns), the actual privacy cost is $2\varepsilon$ even though the system reports $\varepsilon$.

\begin{figure}
\caption{The accounting bug and the data-dependent control flow bug raised in the MOSTLY AI codebase.}
\label{fig:mostly-composition}
\begin{lstlisting}[style=mypython, escapechar=@, captionpos=b]
def analyze_reduce_numeric(...
    value_protection: bool = True,
    value_protection_epsilon: float | None = None,
    ...) -> dict:
    @\colorbox{red!20}{BUG: both mechanisms spend the same privacy budget}@
    @\colorbox{red!20}{... = dp\_approx\_bounds(..., value\_protection\_epsilon)}@
        ...
    @\colorbox{red!20}{... = dp\_non\_rare(..., value\_protection\_epsilon)}@
    ...

def _analyze_reduce_seq_len(....)
    @\colorbox{red!20}{BUG: np.any(lengths != 1) is a non-private measurement}@
    if value_protection_epsilon is not None
        and np.any(lengths != 1):
        ... = dp_quantiles(...)
    # dp_quantiles skipped otherwise
    
\end{lstlisting}
\end{figure}

Additionally, we observed data-dependent branching, where control-flow decisions rely on raw, non-private statistics. For example, the decision to invoke the private quantile routine depends on whether any sequence length exceeds one (\Cref{fig:mostly-composition}). An observer can therefore infer the existence of multi-step sequences solely from which mechanisms are called, as detected by our invariance check.

A similar pattern appears in the numeric encoder, where the unique value count is computed non-privately and compared to a threshold (e.g., $<100$) to decide whether to compute per-value counts. Crossing this alters the mechanisms invoked, revealing if the cardinality lies above or below the boundary.

\subsubsection{Opacus} We audited the \texttt{make\_private} function, which instruments a model for DP training. As shown in \Cref{fig:opacus-bug}, the variable \texttt{expected\_batch\_size} is derived from the private variable \texttt{len(data\_loader.dataset)}. Under unbounded adjacency (assumed by Poisson sampling in the Subsampled Gaussian Mechanism \cite{kasiviswanathan2011can, DBLP:journals/corr/abs-1807-01647, balle_privacy_2020}), the dataset size $n$ is private. Using it for normalization leaks $n$ into the optimizer. While privacy loss decreases as $n$ increases, this constitutes a violation for small datasets, detected via \methodtrusted. Notably, a similar issue was reported over two years ago but remains unpatched \cite{opacus_issue571}. We found no other bugs in their implementation.

\begin{figure}
\caption{Simplified logic from Opacus's \texttt{make\_private} function. The \texttt{sample\_rate} is derived from \texttt{len(data\_loader.dataset)}, which is a private quantity under the add/remove adjacency model, leading to a parameter instability bug.}
\label{fig:opacus-bug}
\begin{lstlisting}[
    style=mypython,       % Use the Python style we defined
    escapechar=@,         % Set '@' as the escape character
    captionpos=b          % Ensure caption is at the bottom
]
def make_private(...data_loader: DataLoader...):
    ...
    sample_rate = 1 / len(data_loader)
    @\colorbox{red!20}{ BUG: size of the dataset is private under add/remove}@
    @\colorbox{red!20}{ expected\_batch\_size = int(len(data\_loader.dataset)}@
                            @\colorbox{red!20}{ * sample\_rate)}@
    ...
\end{lstlisting}
\end{figure}

\subsubsection{dpmm} We audited the \texttt{dpmm} library and uncovered an invariance violation in both the \texttt{AIM} and \texttt{PrivBayes} algorithms via \methodtrusted. When no domain is pre-specified for the marginals, the implementation derives it directly from the private data via
$$\texttt{\_domain = (df.astype(int).max(axis=0) + 1).to\_dict()}$$ without accounting for it in the privacy budget. The presence of a single record with a large feature value can thus change the inferred domain across neighboring datasets, altering the marginals' supports. In \methodtrusted{}, this manifests as an invariance violation between the Record and Replay runs. A robust DP implementation should treat the domain as public (provided by the user or a safe default), or estimate it through an explicit differentially private primitive rather than inferring it directly from the private data.

\subsubsection{Vulnerability to pathological inputs} We also audited the robustness of the implementations against pathological inputs, specifically $\pm \infty$ and \texttt{NaN} values. Correctly handling these inputs under differential privacy is often overlooked, however, they often bypass standard clipping checks (e.g., standard comparison operators with \texttt{NaN} often yield false, effectively skipping bounds enforcement), leading to undefined behavior or information leakage. The results, summarized in \Cref{tab:dp_infnan}, indicate a widespread lack of defensive programming in the Python ecosystem, leading to either unhandled exceptions or the propagation of \texttt{NaN} values into the final output, leaking properties of the input data. Notably, \textit{Google Differential Privacy} is the only library among those tested that consistently handles these inputs correctly, either by filtering them out or raising a controlled error before the privacy budget is consumed.

\subsubsection{Additional evaluations and scope} We also audited OpendDP's Python bindings for MST and AIM \cite{Shoemate_OpenDP_Library}, Google Differential Privacy's Clustering algorithm \cite{chang_locally_nodate}, Jax Privacy \cite{jax-privacy2022github}, Tensorflow Privacy \cite{tensorflow_privacy}, and PipelineDP \cite{openmined_pipelinedp}. Our framework did not detect any privacy violations in these libraries. We note that our evaluation scope is currently restricted to local Python execution models. Consequently, we excluded libraries implemented in other languages or frameworks that rely on distributed computing backends, such as Tumult Analytics \cite{berghel_tumult_2022} and Privacy on Beam \cite{google_privacy_beam}. As noted in \Cref{sec:suggestionsforpractitioners}, adapting our framework to distributed environments represents an important direction for future work.

\begin{table}
    \centering
    \caption{Summary of vulnerabilities detected across various DP libraries. A (green) cross \greenx ~ indicates that our framework did not catch a bug, while a (red) checkmark \redcheck ~ indicates that the framework found a bug.} 
    \label{tab:dp_bugs}
    \scriptsize
    \renewcommand{\arraystretch}{1.3}
    
    \renewcommand\theadfont{\bfseries} 
    
    \renewcommand\tabularxcolumn[1]{m{#1}}
    \newcolumntype{Y}{>{\centering\arraybackslash}X}

    \begin{tabularx}{\columnwidth}{@{} m{3cm} Y Y Y @{}} 
        \toprule
        \thead[l]{Implementation} & 
        \thead{Sensitivity\\Miscalibration} & 
        \thead{Noise\\Miscalibration} & 
        \thead{Invariant\\Violation} \\
        \midrule
        
        \multicolumn{4}{@{}l}{\textit{\textbf{SmartNoise Synth}}} \\
        \hspace{5mm} Logistic Regression & \redcheck & \greenx & \greenx \\

        \addlinespace
        \multicolumn{4}{@{}l}{\textit{\textbf{SmartNoise SQL}}} \\
        \hspace{5mm} Odometers & \redcheck & \greenx & \greenx \\
        
        \addlinespace
        \multicolumn{4}{@{}l}{\textit{\textbf{dpmm}}} \\
        \hspace{5mm} PrivBayes & \greenx & \greenx & \redcheck \\
        \hspace{5mm} MST        & \greenx & \greenx & \greenx \\
        \hspace{5mm} AIM        & \greenx & \greenx & \redcheck \\

        \addlinespace
        \multicolumn{4}{@{}l}{\textit{\textbf{Private PGM}}} \\
        \hspace{5mm} MWEM+PGM & \greenx & \greenx & \greenx \\
        \hspace{5mm} JAM        & \redcheck & \greenx & \greenx \\
        \hspace{5mm} Gaussian+AppGM        & \greenx & \greenx & \greenx \\
        \hspace{5mm} MST        & \greenx & \greenx & \greenx \\
        \hspace{5mm} AIM        & \greenx & \greenx & \greenx \\
        
        \addlinespace
        \multicolumn{4}{@{}l}{\textit{\textbf{Synthcity}}} \\
        \hspace{5mm} PrivBayes & \greenx & \redcheck & \redcheck \\
        \hspace{5mm} AIM        & \greenx & \greenx & \greenx \\
        
        \addlinespace
        \multicolumn{4}{@{}l}{\textit{\textbf{Diffprivlib}}} \\
        \hspace{5mm} Linear Regression & \redcheck & \greenx & \greenx \\
        \hspace{5mm} Logistic Regression         & \greenx & \greenx & \redcheck \\

        \addlinespace
        \multicolumn{4}{@{}l}{\textit{\textbf{MostlyAI}}} \\
        \hspace{5mm} Synthethic Data Engine & \greenx & \redcheck & \redcheck \\

        \addlinespace
        \multicolumn{4}{@{}l}{\textit{\textbf{Opacus}}} \\
        \hspace{5mm} DP-SGD & \greenx & \greenx & \redcheck \\
        
        \addlinespace
        \multicolumn{4}{@{}l}{\textit{\textbf{Jax-Privacy}}} \\
        \hspace{5mm} DP-SGD & \greenx & \greenx & \greenx \\

        \addlinespace
        \multicolumn{4}{@{}l}{\textit{\textbf{Google Differential Privacy}}} \\
        \hspace{5mm} Clustering & \greenx & \greenx & \greenx \\

        \addlinespace
        \multicolumn{4}{@{}l}{\textit{\textbf{PipelineDP}}} \\
        \hspace{5mm} DP Engine & \greenx & \greenx & \greenx \\

        \addlinespace
        \multicolumn{4}{@{}l}{\textit{\textbf{OpenDP}}} \\
        \hspace{5mm} MST & \greenx & \greenx & \greenx \\
        \hspace{5mm} AIM & \greenx & \greenx & \greenx \\

        \bottomrule
    \end{tabularx}
\end{table}

\section{Discussion}
\label{sec:suggestionsforpractitioners}

Our audit of major libraries suggests that ensuring the correctness of mathematical derivations is necessary but not sufficient. While the core privacy mechanisms may be theoretically sound, our findings highlight a distinct class of privacy-impacting bugs that warrant investigation. These violations arise not from the privacy proofs themselves, but from the friction between abstract privacy definitions and practical software engineering. Based on the common failure modes detected by our auditing framework, we offer the following recommendations for developers of DP software.

\paragraph{Prioritize Vetted DSLs} In line with prior work \cite{tramer_debugging_2022, annamalai2024you}, our auditing results demonstrate that ad hoc implementations of differential privacy are notoriously error-prone. Consequently, whenever architectural constraints allow, practitioners should prioritize the use of vetted DSLs such as OpenDP \cite{opendp_opendp_2020} or Tumult Analytics \cite{berghel_tumult_2022}. These frameworks abstract away the complexity of sensitivity tracking and noise calibration, significantly reducing the surface area for the types of implementation bugs we have identified.

\paragraph{Unclear neighboring definitions} A recurring source of issues is the ambiguity regarding the adjacency relationship (i.e., the definition of a neighboring dataset). The lack of clear statements can lead to the composition of incompatible mechanisms or incorrect usage. Maintainers should not assume that users understand whether the system provides add/remove or replace-one privacy. Libraries should explicitly encode the adjacency type in the privacy accountant object. A good example is the Google DP accounting library, which requires selecting a neighboring definition \cite{google_dp_event, doroshenko2022connect}, as does Tumult Analytics (see \cite{tumult_analytics_promise}). If a system operates under replace-one, the code can treat the dataset size $n$ as a public constant. However, under the add/remove model, developers must treat $n$ as private information. Our audit of Opacus (\Cref{fig:opacus-bug}) showed that calculating normalization factors based on \texttt{len(dataset)} is a dangerous default. We recommend requiring users to specify the dataset size explicitly or, if it is not public, privately releasing it (at a privacy cost).

\paragraph{Data-dependent code isolation} We recommend structuring DP algorithms by separating data-dependent and data-independent components. All data-dependent preprocessing (clipping, bounding) should be performed in a distinct phase, and the resulting restricted data should then be passed to an isolated core privacy block that handles aggregation and noise addition. Post-processing can be performed afterward. This separation helps identify any unintended data-dependent dependencies outside the privacy mechanism. Additionally, it simplifies the deployment of our auditing technique, allowing us to instrument functions that access private data separately from those that should remain invariant under a neighboring dataset change.

\subsection{Limitations}
\label{sec:limitations}

\paragraph{On the need for~\Cref{assumption:randomness-controlability}} This assumption is central to our methodology and a standard requirement for deterministic testing and debugging in scientific computing. Our goal is to verify that any two execution paths on $D$ and $D'$ are control-flow invariant, with differences only arising from the sensitive inputs to the DP primitives. Without controlling randomness, this verification is impossible. A stochastic decision (e.g., \texttt{if np.random.rand() > 0.5}) taken after the first DP mechanism call could diverge between the "record" and "replay" runs purely by chance, causing our tool to incorrectly flag a data-dependent control-flow bug. Importantly, DP libraries introduce additional sources of randomness beyond the DP mechanisms themselves—for instance, \cite{chang_locally_nodate} uses locality-sensitive hashing \cite{10.1145/997817.997857}, while Opacus and TensorFlow sample minibatches to compute stochastic gradients.

We stress that \Cref{assumption:randomness-controlability} is practical for the vast majority of DP libraries, which are written in high-level languages like Python or C++ and rely on standard, stateful PRNGs (e.g., \texttt{numpy.random}, \texttt{torch.manual\_seed}). We however acknowledge two limitations. First, our method cannot audit systems that draw randomness from non-deterministic sources like \texttt{/dev/random} or specialized cryptographic hardware (e.g., Intel RDRAND), unless these sources can be mocked or patched at a lower level. Second, its application to highly concurrent or distributed systems (such as Tumult Analytics \cite{berghel_tumult_2022} that relies on Apache Spark \cite{zaharia_apache_2016}) is non-trivial, as it would require controlling and synchronizing the PRNG state across multiple threads or processes, which is outside our current scope.

\paragraph{What does our audit mean?} It is important to interpret the results of our audit within the appropriate context. Our \methodtrusted{} framework is a testing tool, not a formal verifier. This distinction introduces a fundamental trade-off: the methodology is sound but not complete:
\begin{enumerate}
    \item \textbf{Soundness (Structural Correctness):} If our tool flags a bug, it corresponds to a concrete implementation error. While it is theoretically possible to construct a data-dependent computation outside privacy mechanisms that does not alter the final output distribution (e.g., a redundant operation), such patterns violate the principle of data-independence outside trusted primitives. Thus, the tool provides a proof-of-failure regarding the \emph{structural} integrity of the algorithm.
    \item \textbf{Incompleteness (False Negatives):} A successful audit (i.e., finding no bugs) does not constitute a proof of correctness. It only indicates that, for the specific neighboring datasets $(D, D')$ and the execution paths tested, no privacy violations were detected.
\end{enumerate}

This limitation, inherent to empirical auditing, is a manifestation of the test coverage problem. A bug may go undetected if it only appears on an untested execution path (i.e., triggered by a dataset $D$ with specific properties) or if it depends on a specific relationship between $D$ and $D'$ not included in the test suite. Additionally, bugs may be missed if they arise from rare stochastic events, such as low-probability failures (e.g., $0.01\%$). Random sampling might not trigger the specific conditions required to observe the failure. Thus, while our tool detects common and critical bugs, it cannot prove the absence of all bugs.

\paragraph{Interpreting the errors.} While our framework offers much higher granularity than black-box auditing by pinpointing the specific primitive call where a violation occurs, it does not automatically reveal the root cause. For instance, if \methodtrusted{} flags a sensitivity violation, it conclusively shows that the inputs to a mechanism exceeded the declared bounds, but it cannot explain \emph{why} the preprocessing logic failed to enforce them. The error might result from a missing clipping step or an incorrect variable reference. Our tool provides the necessary evidence, but identifying the underlying flaw in the source code remains a manual debugging task for the developer.

\subsection{Future work}
\label{sec:futurework}

\paragraph{Extending to other types of accounting and composition}
Currently, our framework assumes adaptive composition \cite{dwork_boosting_2010}, flattening the algorithm's structure into a sequence of mechanism calls. While sufficient for basic sequential composition, this view obscures structural dependencies required for other accounting techniques. For example, the trace alone does not reveal whether two mechanisms operate on disjoint data subsets, which would permit \emph{parallel composition} \cite{dwork_algorithmic_2013}. Similarly, advanced methods such as privacy amplification by iteration \cite{feldman_privacy_2018} rely on the algorithm's semantic structure and the precise sequence of mechanisms. Future work aims to enhance the instrumentation to support these more complex composition techniques.

\paragraph{On the need for \texttt{ensure\_equality} nodes}
Currently, the placement of these verification nodes requires manual annotation: developers must explicitly identify which variables (e.g., hyperparameters, loop counters, configuration settings) should remain invariant. Although effective, this process is susceptible to human oversight. In future iterations, we aim to automate this discovery through advanced dynamic instrumentation, enabling the framework to infer and monitor public invariants automatically. This would also reduce modifications to the audited codebase while maintaining robust coverage against data-dependent control-flow leaks.

\paragraph{Advanced dataset crafting} While our current framework relies on basic dataset crafting strategies, we aim to extend the test suite to support advanced dataset crafting techniques (see for example \cite{9519405}) to enhance bug discovery and testing coverage.

\section{Conclusion}

In this work, we introduce a  novel “gray-box” auditing paradigm that narrows the gap between theoretical claims and the practical guarantees of DP libraries. By shifting the focus from computationally intensive output analysis to precise internal-state inspection, \methodtrusted{} enables developers to verify sensitivity claims and detect data-dependent behavior with the speed and reliability of a unit test. For scenarios involving unverified components, \methoduntrusted{} bridges deterministic checking and statistical validation, providing a fine-grained assessment of complex pipelines. The discovery of real bugs in major DP libraries highlights the need for improved tooling in the privacy ecosystem. By releasing our framework as an open-source package, we hope to help practitioners to build privacy software that faithfully fulfills its theoretical promises.

\section{Ethics and Responsible Disclosure}

During our evaluation of our testing framework, we identified privacy vulnerabilities in existing open-source differential privacy libraries. We followed standard responsible disclosure protocols by reporting these findings to the respective library maintainers. We have established a 60-day disclosure embargo to allow sufficient time for patches to be implemented before the details of these vulnerabilities are made public. We will update the text to include information on how the bugs were addressed.

All experiments were conducted locally using purely synthetic data.

\begin{acks}
This work is partially supported by grant ANR-20-CE23-0015 (Project PRIDE), ANR 22-PECY-0002 IPOP (Interdisciplinary Project on Privacy) project of the Cybersecurity PEPR. This work was performed using HPC resources from GENCI–IDRIS (Grant 2023-AD011014018R3).
\end{acks}

\bibliographystyle{ACM-Reference-Format}
\bibliography{main}

\appendix
\newpage
\section{General Auditing Algorithm}
\Cref{alg:audit-pipeline} describes the general statistical process for privacy auditing.

\begin{algorithm}[h]
\caption{General Privacy Auditing Pipeline}
\label{alg:audit-pipeline}
\begin{algorithmic}[1]
\Statex \textbf{Input:} Mechanism $\mathcal{M}$, Adversary $\mathcal{A}$ (with \texttt{Rank}, \texttt{Reject}), base dataset $D$, canary $x^*$, number of runs $R$, target $\delta$
\Statex \textbf{Output:} Empirical privacy lower bound $\hat{\varepsilon}$

\State $D_0 \leftarrow D$
\State $D_1 \leftarrow D \cup \{x^*\}$
\State $S \leftarrow []$ \Comment{List to store (score, ground\_truth) pairs}

\For{$i = 1$ to $R$}
    \State $b_i \leftarrow \text{Bernoulli}(0.5)$ \Comment{Draw random bit for ground truth}
    \If{$b_i = 0$}
        \State $o_i \leftarrow \mathcal{M}(D_0)$
    \Else
        \State $o_i \leftarrow \mathcal{M}(D_1)$
    \EndIf
    \State $s_i \leftarrow \mathcal{A}$.\Call{Rank}{$o_i, D_0, D_1$} \Comment{Get confidence score}
    \State $S.\text{add}((s_i, b_i))$
\EndFor

\State $b_{pred} \leftarrow \mathcal{A}$.\Call{Reject}{$S, h^*$}
\State (TN, FP, FN, TP) $\leftarrow$ \Call{ComputeStatistics}{$S.\text{labels}, b_{pred}$}
\State $(\alpha, \beta) \leftarrow$ \Call{ConfInterval}{TN, FP, FN, TP} \Comment{e.g., CP \cite{jagielski2020auditingdifferentiallyprivatemachine}}
\State $\hat{\varepsilon} \leftarrow$ \Call{ConvertToDP}{$\alpha, \beta, \delta$} \Comment{e.g., via GDP \cite{nasr2023tight}}

\State \textbf{return} $\hat{\varepsilon}$
\end{algorithmic}
\end{algorithm}

\section{Details on Empirical PLD Estimation}
\label{appendix:pld_estimation}

In \methoduntrusted, we aim to integrate black-box statistical auditing for \emph{untrusted} privacy mechanisms, enabling verification of the actual privacy loss. Unlike other auditing techniques that either evaluate a single $(\varepsilon, \delta)$ (e.g. \cite{nasr_adversary_2021, jagielski2020auditingdifferentiallyprivatemachine}), or that rely on the structure of the underlying mechanism (e.g. \cite{nasr2023tight}) \emph{empirical Privacy Loss Distribution (PLD)}. This allows us to compose empirically audited mechanisms with theoretically trusted ones using standard numerical accountants~\cite{doroshenko2022connect}, and to obtain the overall privacy loss for the pipeline for some fixed $\delta$.

Our techniques uses trade-off function, for a clear overview please see \cite{dong2022gaussian} and PLDs (see \cite{dwork_concentrated_2016, doroshenko2022connect}. The estimation procedure consists of three steps: (i) estimating the trade-off function via binary classification, (ii) converting the trade-off to the privacy profile $(\varepsilon, \delta(\varepsilon))$, and (iii) reconstructing a pessimistically discretized PLD using the privacy profile.  
\paragraph{Trade-off Function Estimation}
Let $\mathcal{M}$ be a mechanism and $D, D'$ be a pair of adjacent datasets. We denote the output distributions as $P = \mathcal{M}(D)$ and $Q = \mathcal{M}(D')$. To empirically estimate the privacy leakage, we interpret the distinguishing problem as a hypothesis testing task between $H_0: o \sim P$ and $H_1: o \sim Q$. We generate empirical sets of output samples $S_P = \{o_i\}_{i=1}^n$ drawn from $\mathcal{M}(D)$ and $S_Q = \{o'_i\}_{i=1}^n$ drawn from $\mathcal{M}(D')$.

We train a binary classifier $\mathcal{C}$ (e.g., via logistic regression) to discriminate between $S_P$ and $S_Q$. The performance of $\mathcal{C}$ serves as a proxy for the optimal decision rule. We compute the Receiver Operating Characteristic (ROC) curve, yielding a set of error pairs $\{(\alpha_k, \beta_k)\}_k$. Here, $\alpha$ represents the Type I error (False Positive Rate) and $\beta$ represents the Type II error (False Negative Rate). These pairs approximate the true trade-off function $f(\alpha)$, formally defined as the minimum Type II error achievable for a given Type I error constraint $\alpha$:

\begin{equation}
    f(\alpha) = \inf_{\phi} \{ 1 - \mathbb{E}_{o \sim Q}[\phi(o)] \mid \mathbb{E}_{o \sim P}[\phi(o)] \leq \alpha \},
\end{equation}

\noindent where $\phi$ represents a rejection rule $0 \leq \phi(\cdot) \leq 1$.

\paragraph{Conversion to the Privacy Profile} With the tradeoff function $f$ at hand, we rely on the duality between $f$-DP and $(\varepsilon, \delta)$-DP established by ~\citet{dong2022gaussian}, Proposition 2.12. The privacy profile $\delta(\varepsilon)$ is the convex conjugate of the trade-off function $f$. We compute the empirical privacy profile $\delta(\varepsilon)$ for a grid of $\varepsilon$ values by solving the following optimization:
\begin{equation}
    \delta(\varepsilon) = 1 - \inf_{\alpha \in [0,1]} \left( e^\varepsilon \alpha + f(\alpha) \right).
\end{equation}
In our implementation, we interpolate the discrete empirical pairs $\{(\alpha_k, \beta_k)\}$ to approximate $f(\alpha)$ and solve the minimization problem numerically using \texttt{scipy.optimize}.

\paragraph{PLD Reconstruction} Finally, to enable composition with other mechanisms, we convert the empirical profile $\delta(\varepsilon)$ into a PLD. For accounting purposes, we require a PLD that \emph{upper bounds} the privacy loss. We construct a discretized PLD defined over a grid of privacy loss values $x_k = k \cdot \Delta x$. For each $x_k$, we determine the maximum probability mass consistent with the empirical $\delta(\varepsilon)$ curve. Specifically, we employ a ``connect-the-dots'' strategy that constructs a pessimistic Probability Mass Function (PMF) such that the $\delta(\varepsilon)$ implied by the PMF is always greater than or equal to the empirically measured $\delta(\varepsilon)$. This empirical PLD is then composed with the PLDs of trusted components to derive the final audit result. To do this, we have used the \texttt{create\_pmf\_pessimistic\_connect\_dots} of \citet{doroshenko2022connect} in the Google DP library.

\section{Audited Codebases}
\label{appendix:audited_codebases}

\paragraph{SmartNoise SDK \cite{noauthor_smartnoise_nodate}} Developed jointly by Microsoft and the OpenDP \cite{opendp_opendp_2020} project, SmartNoise is a widely deployed platform designed for general-purpose analytics. It focuses on tabular data protection and SQL query rewriting. We specifically audited its core synthesizer components and covariance release mechanisms.

\paragraph{dpmm \cite{mahiou_dpmm_2025}} The \texttt{dpmm} library is a specialized toolkit for generating differentially private synthetic data using marginal models. It implements state-of-the-art algorithms such as PrivBayes \cite{10.1145/3134428}, MST \cite{mckenna_winning_2021}, and AIM \cite{mckenna_aim_2024}. Unlike general-purpose libraries, \texttt{dpmm} focuses exclusively on preserving the statistical utility of low-dimensional marginals, making it a good candidate for testing the robustness of complex, iterative sampling mechanisms.

\paragraph{Synthcity \cite{qian_synthcity_2023}} Synthcity is a comprehensive framework for synthetic data generation. It includes a vast collection of generative models, ranging from GANs to Bayesian networks. For this audit, we targeted the privacy-preserving modules to verify whether the diverse set of integrated algorithms consistently adheres to their claimed privacy budgets, including their variant of PrivBayes and AIM.

\paragraph{Opacus \cite{yousefpour_opacus_2022}} Maintained by Meta, Opacus is the de facto standard library for training PyTorch \cite{paszke2019pytorch} models with DP. It implements Differentially Private Stochastic Gradient Descent (DP-SGD) \cite{abadi_deep_2016} by hooking into the optimizer step to perform per-sample gradient clipping and noise addition. Given its dominance in deep learning, ensuring the correctness of its gradient computation logic is an exciting task for our auditing framework.

\paragraph{Diffprivlib \cite{holohan_diffprivlib_2019}} IBM's DP Library is a general-purpose toolkit designed to mimic the API of the popular \texttt{scikit-learn} library. It provides drop-in replacements for standard machine learning algorithms (e.g., Logistic Regression, Naive Bayes) and pre-processing tools. We selected it to test how standard ML paradigms are adapted for DP and whether the abstraction layers introduce errors.

\paragraph{Jax-privacy  \cite{jax-privacy2022github}} Developed by Google DeepMind, \texttt{jax-privacy} leverages the JAX \cite{jax2018github} high-performance computing framework to enable DP training for large-scale models. It is designed for research flexibility and performance, often serving as the reference implementation for new privacy-preserving optimization techniques. We audited this library to evaluate privacy enforcement in a functional programming paradigm, which differs significantly from the object-oriented approach of PyTorch-based libraries.

\paragraph{Private-PGM} Next, we turn our attention to the framework widely known for hosting the winning submission of the NIST US-UK PETS Prize, namely Private-PGM (\cite{mckenna2021winning}). In the package a number of synthetic data models are defined, including the JAM mechanism. 

\paragraph{MostlyAI \cite{mostlyai}} MostlyAI offers an enterprise-grade platform for high-fidelity synthetic data generation. We examined their open-source Python client, which facilitates the training of generative models and the generation of synthetic datasets. Our audit focused on the client-side implementation of privacy safeguards and the mechanisms used to interface with their generative engine.

\paragraph{Google Differential Privacy \cite{chang_locally_nodate}} This repository contains the foundational building blocks for differential privacy, with implementations in C++, Go, and Java. It provides the core algorithms (e.g., Laplace, Gaussian, and Count mechanisms) that power many higher-level tools. We have only audited their clustering mechanism \cite{chang_locally_nodate}, which is fully implemented in Python.

\paragraph{PipelineDP \cite{openmined_pipelinedp}} PipelineDP is a Python framework developed by OpenMined and Google for applying differentially private aggregations to large datasets.

\paragraph{SmartNoise SQL \cite{noauthor_smartnoise_nodate}} Distinct from the analysis SDK, SmartNoise SQL serves as the relational database interface for the SmartNoise project. It operates by intercepting standard SQL queries and rewriting them into differentially private equivalents compatible with backend engines like PostgreSQL. We specifically analyzed the \texttt{Odometer} mechanism, which tracks, accumulates, and enforces privacy budget limits across sequences of queries.

\paragraph{OpenDP \cite{Shoemate_OpenDP_Library}} OpenDP is a community-governed project designed to serve as a trustworthy, verifiable kernel of algorithms for differential privacy. Built primarily in Rust to leverage its strict memory safety and type systems, the library also enables Python-based implementations of complex algorithms like MST and AIM. We audited these mechanisms to verify that they correctly compose OpenDP's foundational primitives (i.e. transformations and measurements) to maintain privacy guarantees.

\paragraph{Reproducibility and Scope} Table \ref{tab:libraries} clarifies the specific software artifacts examined in this paper. Given the rapid development of open-source DP tools, reproducibility is essential. To this end, we report the exact commit hashes for the code snapshots we audited. 

\begin{table}[ht]
    \centering
    \scriptsize 
    \setlength{\tabcolsep}{2pt} 
    
    \caption{Overview of the audited libraries, with associated versions and targeted modules for reproducibility purposes.}
    \label{tab:libraries}
    
    \begin{tabularx}{\columnwidth}{@{} l l l X @{}}
        \toprule
        \textbf{Library} & \textbf{Link} & \textbf{Commit Hash} & \textbf{Relevant Files \& Functionality} \\
        \midrule
        
        SmartNoise \cite{noauthor_smartnoise_nodate} & 
        \href{https://github.com/opendp/smartnoise-sdk/commit/b4f0058d27eb08b0784864170a1c0bfa875b1c6f}{GitHub} & 
        b4f0058 & 
        \href{https://github.com/opendp/smartnoise-sdk/blob/b4f0058d27eb08b0784864170a1c0bfa875b1c6f/synth/snsynth/models/dp_covariance.py}{Covariance Estimator} \\
        \addlinespace
        
        dpmm \cite{mahiou_dpmm_2025} & 
        \href{https://github.com/sassoftware/dpmm/commit/e5cd92bfc02a9236c4b7e504bb357076c3d4423e}{GitHub} & 
        e5cd92b & 
        \href{https://github.com/sassoftware/dpmm/blob/e5cd92bfc02a9236c4b7e504bb357076c3d4423e/src/dpmm/models/base/mechanisms/mechanism.py\#L113}{AIM Mechanism Domain},
        \href{https://github.com/sassoftware/dpmm/blob/e5cd92bfc02a9236c4b7e504bb357076c3d4423e/src/dpmm/models/priv_bayes.py}{PrivBayes}
        \\
        \addlinespace
        
        Synthcity \cite{qian_synthcity_2023} & 
        \href{https://github.com/vanderschaarlab/synthcity/tree/23f322fe381326ed01c41b13d469a06e38cce545}{GitHub} & 
        23f322f & 
         \href{https://github.com/vanderschaarlab/synthcity/blob/23f322fe381326ed01c41b13d469a06e38cce545/src/synthcity/plugins/privacy/plugin_privbayes.py}{PrivBayes} \\
        \addlinespace
        
        Opacus \cite{yousefpour_opacus_2022} & 
        \href{https://github.com/meta-pytorch/opacus/tree/f17f254ab8f1f1095e8257bf278769d549748bbc}{GitHub} & 
        f17f254 & 
         \href{https://github.com/meta-pytorch/opacus/blob/f17f254ab8f1f1095e8257bf278769d549748bbc/opacus/privacy_engine.py}{\texttt{make\_private}}\\
        \addlinespace
        
        Diffprivlib \cite{holohan_diffprivlib_2019} & 
        \href{https://github.com/IBM/differential-privacy-library/tree/5d9c9d873c99295f14b0cfb61866fafbf3cc4684}{GitHub} & 
        5d9c9d8 & 
        \href{https://github.com/IBM/differential-privacy-library/blob/5d9c9d873c99295f14b0cfb61866fafbf3cc4684/diffprivlib/models/linear_regression.py}{Linear Regression} \\
        \addlinespace
        
        Jax-privacy \cite{jax-privacy2022github} & 
        \href{https://github.com/google-deepmind/jax_privacy/tree/2ddd6d93aa21140823d70f7e75ad6186624ebc75}{GitHub} & 
        2ddd6d9 & 
        - (No bugs found) \\
        \addlinespace
        
        Private-PGM \cite{mckenna2021winning} & 
        \href{https://github.com/ryan112358/mbi/tree/802218260d5684452ac4a9c44bc08cf2b2354d4b}{GitHub} & 
        8022182 & 
        \href{https://github.com/ryan112358/mbi/blob/802218260d5684452ac4a9c44bc08cf2b2354d4b/mechanisms/jam.py\#L108}{JAM} \\
        \addlinespace

        MostlyAI \cite{mostlyai} & 
        \href{https://github.com/mostly-ai/mostlyai-engine/tree/00cfb12efccb24099668c1b97f8dcd5ca9071960}{GitHub} & 
        00cfb12 & 
        \href{https://github.com/mostly-ai/mostlyai-engine/blob/00cfb12efccb24099668c1b97f8dcd5ca9071960/mostlyai/engine/_encoding_types/tabular/numeric.py\#L233}{analyze\_reduce\_numeric} \\
        \addlinespace
        
        
        SmartNoise SQL \cite{noauthor_smartnoise_nodate} & 
        \href{https://github.com/opendp/smartnoise-sdk/tree/81d1351b383b5d45b0818ae140a7291b4ce4a841/sql}{GitHub} & 
        81d1351 & 
        \href{https://github.com/opendp/smartnoise-sdk/blob/81d1351b383b5d45b0818ae140a7291b4ce4a841/sql/snsql/sql/odometer.py}{Odometer} \\
        \addlinespace
        
        Google Differential Privacy \cite{chang_locally_nodate} & 
        \href{https://github.com/google/differential-privacy/tree/3cff255e520d1c446025fc7030cf549d8659ddab}{GitHub} & 
        3cff255 & 
        - (No bugs found) \\
        \addlinespace
        
        PipelineDP \cite{openmined_pipelinedp} & 
        \href{https://github.com/OpenMined/PipelineDP}{GitHub} & 
        94bacde  & 
        - (No bugs found) \\
        \addlinespace

        OpenDP \cite{Shoemate_OpenDP_Library} & 
        \href{https://github.com/opendp/opendp}{GitHub} & 
        cc3a18b  & 
        - (No bugs found) \\
        \addlinespace
        
        \bottomrule
    \end{tabularx}
\end{table}

\section{Handling of Adversarial Inputs}

Table~\ref{tab:dp_infnan} demonstrates the effect of using adversarial inputs containing NaN values and infinity. While these values were likely out of scope for the original framework, we showcase the widespread implication of not performing tight input validation on mechanisms. 

\begin{table}[t]
    \centering
    \caption{Summary of behavior under $-\infty$ and NaN inputs across the evaluated DP libraries. A (green) cross \greenx{} indicates that the implementation correctly handled the corresponding case, while a (red) checkmark \redcheck{} indicates that our framework found an issue when injecting $\pm \infty$/NaN values.}
    \label{tab:dp_infnan}
    \scriptsize
    \renewcommand{\arraystretch}{1.3}
    \newcolumntype{Y}{>{\centering\arraybackslash}X}

    \begin{tabularx}{\columnwidth}{@{} l Y Y @{}}
        \toprule
        \textbf{Implementation} & 
        \thead{\textbf{$\pm \infty$}\\\textbf{Inputs}} & 
        \thead{\textbf{NaN}\\\textbf{Inputs}} \\
        \midrule
        
        \multicolumn{3}{@{}l}{\textit{\textbf{SmartNoise Synth}}} \\
        \hspace{5mm} Logistic Regression & \redcheck & \redcheck \\

        \addlinespace
        \multicolumn{3}{@{}l}{\textit{\textbf{dpmm}}} \\
        \hspace{5mm} PrivBayes & \redcheck & \redcheck \\
        \hspace{5mm} MST       & \greenx & \greenx \\
        \hspace{5mm} AIM       & \redcheck & \redcheck \\

        \multicolumn{3}{@{}l}{\textit{\textbf{Synthcity}}} \\
        \hspace{5mm} PrivBayes & \redcheck & \redcheck \\
        \hspace{5mm} AIM       & \greenx & \greenx \\
        
        \addlinespace
        \multicolumn{3}{@{}l}{\textit{\textbf{Diffprivlib}}} \\
        \hspace{5mm} Linear Regression & \redcheck & \redcheck \\
        \hspace{5mm} Logistic Regression  & \redcheck & \redcheck \\

        \addlinespace
        \multicolumn{3}{@{}l}{\textit{\textbf{Opacus}}} \\
        \hspace{5mm} DP-SGD & \redcheck & \redcheck \\

        \addlinespace
        \multicolumn{3}{@{}l}{\textit{\textbf{Google Differential Privacy}}} \\
        \hspace{5mm} Clustering & \greenx & \greenx \\

        \bottomrule
    \end{tabularx}
\end{table}

\section{Acknowledgments}
The authors used generative AI-based tools to revise the text, improve flow, and correct typos, grammatical errors, and awkward phrasing. AI-based tools were used in the production of the code as well for documentation writing, test writing, polishing, and for the understanding and discovery of the audited codebases.

\end{document}